\documentclass[superscriptaddress,a4paper,nofootinbib,amsmath,amssymb,floatfix,prd]{revtex4-2}

\usepackage{array} 
\usepackage{booktabs}
\usepackage{orcidlink}
\usepackage{amsmath}
\usepackage{amssymb}
\usepackage{siunitx}  
\usepackage{multirow} 
\usepackage{rotating}   
\usepackage{tabularx}   
\usepackage{comment}
\usepackage{ulem} 
\usepackage{graphicx}
\usepackage{xcolor}
\usepackage{color}
\usepackage{caption}
\usepackage{subcaption}
\usepackage{amsmath}
\usepackage{graphicx}
\usepackage{float}

\usepackage{chngcntr}
\numberwithin{equation}{section}
\counterwithin{figure}{section}
\counterwithin{table}{section}

\def\RS{\textcolor{blue}}
\def\AS{\textcolor{cyan}}

\begin{document}

\begin{flushright}
BONN-TH-2026-11
\end{flushright}

\vspace*{0.5cm}


\title{A new approach to long-lived particle detection at hadron colliders: \\the \textsf{DELIGHT-SHIELD} concept}



\author{Biplob Bhattacherjee\,\orcidlink{0000-0003-3668-8305}}
\email{biplob@iisc.ac.in}
\affiliation{Centre for High Energy Physics, Indian Institute of Science, Bangalore 560012, India}

\author{Arnav Chauhan\,\orcidlink{000}}
\email{arnav.chauhan@cern.ch}
\affiliation{Department of Physics, Indian Institute of Technology, Kanpur 208016, India} 
\affiliation{School of Physics and Institute for Collider Particle Physics, University of the Witwatersrand, Johannesburg, Wits 2050, South Africa}  

\author{Swagata Mukherjee\,\orcidlink{0000-0001-6341-9982}} 
\email{swagata@iitk.ac.in}
\affiliation{Department of Physics, Indian Institute of Technology, Kanpur 208016, India} 

\author{\\Rhitaja Sengupta\,\orcidlink{0000-0003-2293-8684}} 
\email{rsengupt@uni-bonn.de}
\affiliation{Bethe Center for Theoretical Physics and Physikalisches Institut der Universit\"at Bonn, \\
Nu\ss allee 12, Bonn 53115, Germany}

\author{Anand Sharma\,\orcidlink{0009-0005-5015-7596}}  
\email{anandsharma2@iisc.ac.in}
\affiliation{Centre for High Energy Physics, Indian Institute of Science, Bangalore 560012, India}

\begin{abstract}
We propose a fundamental shift in the search for beyond the Standard Model long-lived particles (LLPs) at high-luminosity hadron colliders by prioritizing physical background suppression over traditional inner tracking. We introduce \textsf{DELIGHT-SHIELD}, a dedicated detector design for a 100\,TeV Future Circular Collider at a dedicated interaction point for LLP searches. By replacing the inner parts of the detector with a multi-layered composite shield, followed by tracking volumes, we estimate a suppression of Standard Model hadronic and electromagnetic backgrounds by up to seven orders of magnitude analytically. 
Full \textsc{Geant4} simulations validate the effectiveness of this design. Although the achieved suppression is somewhat lower than the analytical estimate, primarily due to secondary particle production within the shield, the residual background remains at a level that is manageable for LLP analyses. It can be further mitigated by applying energy thresholds, as well as vertexing and timing cuts in the downstream detector.
Benchmarking against dark scalar model, we show that this shielding based detector concept achieves sensitivity to branching ratios as low as $\mathcal{O}(10^{-9})$ for $h\rightarrow\phi\phi$ process under zero background condition -- outperforming general-purpose detector baselines. This strategy not only expands the discovery reach for neutral LLPs but also provides a rigorous experimental handle to distinguish new physics from Standard Model punch-through backgrounds. We further discuss a phased implementation at the High-Luminosity LHC as a critical testbed for this novel detection concept.
\end{abstract}

\maketitle

\vspace*{-0.5cm}

\section{Introduction}
\label{sec:intro}

With the Large Hadron Collider (LHC) nearing the end of Run-3 and preparing for a long shutdown for the Phase-II upgrade to the high-luminosity LHC (HL-LHC), it is timely to assess how best to maximize the physics potential of future collider experiments.
The future hadron colliders are all high luminosity machines, having very high rates of pile-up (PU) interactions\,\cite{ZurbanoFernandez:2020cco,FCC:2025lpp}.
As a result, the Standard Model (SM) backgrounds
are expected to adversely affect the discovery potential for long-lived particles (LLPs) at the HL-LHC and the proposed future circular hadron collider (FCC-hh). 
A high PU environment leads to an exceptionally busy detector, driven by the abundance of soft-QCD interactions. This significantly complicates searches for light and weakly coupled new physics signals, which can be easily lost amid the
overwhelming background activity. Moreover, regions of the detector that are typically considered relatively quiet, such as the muon spectrometer (MS), are no longer immune. At a 100\,TeV collider, the increased production of high-energy jets raises the likelihood of calorimeter punch-through, resulting in additional activity in the MS and further degrading its performance.

Motivated from various BSM models\,(see Ref.\,\cite{Alimena:2019zri} for an overview), a multitude of LLP signatures have been explored in several phenomenological studies, such as Refs.\,\cite{deVries:2015mfw, Banerjee:2017hmw, Dercks:2018wum, Bhattacherjee:2019fpt, Banerjee:2019ktv, DeVries:2020jbs, Bhattacherjee:2020nno, Bhattacherjee:2021rml, Bhattacherjee:2021qaa, Adhikary:2022pni, Ovchynnikov:2022its, Bandyopadhyay:2022mej, Bandyopadhyay:2023joz, Bhattacherjee:2023plj, Bhattacherjee:2023evs, Bhattacherjee:2023kxw, Bandyopadhyay:2023lvo, Gunther:2023vmz, deVries:2024mla, Wang:2024ieo,Kim:2025tuz,Bernal:2025qkj,Cheung:2025kmc,Liu:2025bbc,Chen:2025btv,Das:2026qwe} and collider experiments\,\cite{ATLAS:2015xit,ATLAS:2018rjc,ATLAS:2018niw,ATLAS:2018tup,ATLAS:2019fwx,ATLAS:2019tkk,ATLAS:2019jcm,ATL-PHYS-PUB-2019-002,ATLAS:2020xyo,ATLAS-CONF-2021-032,ATLAS:2021jig,CMS:2014hka,CMS:2017kku,CMS:2018bvr,CMS-PAS-FTR-18-002,CMS:2019zxa,CMS:2020atg,CMS-PAS-EXO-19-021,CMS:2021juv,CMS:2021kdm,CMS:2021yhb,LHCb:2016buh,LHCb:2016inz,LHCb:2016awg,LHCb:2017xxn,LHCb:2019vmc,LHCb:2020akw}. Apart from the collider detectors, several beam dump experiments also contribute to LLP searches\,\cite{E949:2008btt,BNL-E949:2009dza,NA62:2020pwi,NA62:2020xlg,NA62:2021zjw,Gorbunov:2021ccu,CHARM:1985anb,Egana-Ugrinovic:2019wzj}. Additionally, there are several proposals for dedicated detectors for neutral LLPs, like  
FASER-2 \cite{Feng:2022inv}, FACET
\cite{Cerci:2021nlb}, MAPP-MoEDAL \cite{MoEDAL-MAPP:2022kyr}, MATHUSLA \cite{Curtin:2018mvb,MATHUSLA:2019qpy,Curtin:2023skh}, CODEX-b \cite{Aielli:2019ivi}, ANUBIS \cite{Bauer:2019vqk}, which are proposed to be placed around different collider interaction points (IP) at 
the High Luminosity LHC (HL-LHC) in the forward or transverse directions.
There are also multiple proposals for dedicated LLP detectors at future collider facilities\,\cite{Blondel:2022qqo,Wang:2019xvx,Chrzaszcz:2020emg,Bhattacherjee:2021rml,Schafer:2022shi,Boyarsky:2022epg,Bhattacherjee:2023plj,MammenAbraham:2024gun,Lu:2024fxs}.

While the four planned interaction points (IPs) of the FCC-hh include two general-purpose detectors\,\cite{FCC:2025lpp}, the remaining IPs offer a unique opportunity to host specialized experiments tailored for weakly-interacting BSM physics. 
To enhance the LLP program at the FCC-hh, we propose a dedicated interaction point with an integrated luminosity of $30\text{ ab}^{-1}$, equipped with a non-conventional detector.
Although conventional inner tracking is essential for prompt physics, we show that sensitivity to neutral LLPs with displaced decay signatures can be enhanced by orders of magnitude by trading inner detector parts for strategic, high-density shielding, followed by a tracking volume. Our central theme is a transition from electronic background rejection to physical background suppression. 
By integrating shielding with immediately downstream tracking layers, hadronic backgrounds can be both attenuated and actively rejected, enabling operation in an effectively near-zero-background regime and allowing for significantly lower trigger thresholds.

To this end, we evaluate four implementation pathways at the FCC-hh. Initially, a standalone detector optimized for the entire 30\,$\text{ab}^{-1}$ run could be set up as a dedicated LLP facility. Alternatively, a phased operation strategy could be adopted in the event that a dedicated IP is not available. This makes use of a modular detector design that allows for the removal of inner trackers via a rail system, enabling an initial LLP-specific configuration ($1\text{--}5\text{ ab}^{-1}$) with the shielding as the inner detector, followed by the tracker, before returning to the baseline general-purpose detector setup. We want to investigate whether the shielding based detector taking data with reduced integrated luminosity runs at the FCC-hh can compete with FCC-hh main detector's Barrel MS in full $30\text{ ab}^{-1}$ luminosity run. 
These shielding-heavy configurations offer a rigorous method to cross-check whether anomalies in the baseline FCC-hh barrel MS or hadron calorimeter (HCAL) are indicative of true BSM physics 
or simply SM background. 
We also explore whether, in cases where the full desired shielding to reduce the SM backgrounds is not feasible, if a thinner shielding configuration can still be implemented to enhance LLP searches in the electromagnetic calorimeter (ECAL).

This shielding-centric approach is further explored as a practical strategy for the HL-LHC. If no evidence of BSM physics is discovered toward the end of the program, unconventional prototyping runs could be initiated to probe previously inaccessible parameter spaces.
We investigate the addition of thin composite layers to stop soft-QCD photons, while acknowledging that thicker shielding is required to suppress high-energy pions. 
Beyond immediate sensitivity, utilizing the HL-LHC in this manner provides a critical testbed for radiation hardness and material performance under a 14\,TeV beam, laying the experimental groundwork for future FCC-hh detector designs.

The rest of the paper is organized as follows. In Section\,\ref{sec:benchmark}, we review the benchmark models chosen to illustrate the sensitivities of the shielding-based detection concept. In Section\,\ref{sec:dedicatedIP}, we introduce and describe the \textsf{DELIGHT-SHIELD} concept at a dedicated interaction point at the FCC-hh, along with the details of the layered shielding required and its sensitivity to the physics benchmarks. In case a dedicated IP for LLP searches is not feasible, we study a few alternative strategies to use the shielding-based search at a general IP of FCC-hh to gain sensitivity for LLPs in Section\,\ref{sec:no_dedicatedIP}. Further, in Section\,\ref{sec:HL_LHC}, we discuss how the concept can be tested and used at the HL-LHC. Finally, we conclude with a summary and outlook in Section\,\ref{sec:summary}.

\section{Benchmark Models}
\label{sec:benchmark}

To evaluate the discovery reach of the \textsf{DELIGHT-SHIELD} concept, we consider two well-motivated extensions of the SM that predict LLPs -- the dark scalar and the heavy neutral lepton (HNL) models. 
These scenarios are motivated by various extensions of the SM that address the dark matter and neutrino mass problems (for example, see Refs.\,\cite{Matsumoto:2018acr,Arcadi:2019lka,Abdullahi:2022jlv}), and they are recommended by the Physics Beyond Colliders (PBC) collaboration\,\cite{Alemany:2019vsk,PBC:2025sny} as representative benchmarks for the Higgs-portal and fermion-portal new physics, respectively.

\vspace*{-0.4cm}

\subsection{Dark Scalar}
\label{ssec:dark_scalar}

\vspace*{-0.3cm}

The dark scalar model introduces a real singlet scalar, $\phi$, that interacts with the SM via the Higgs portal~\cite{Ferber:2023iso}. This interaction is characterized by a mixing angle, $\theta$, whereby $\phi$ inherits the SM Higgs couplings scaled by $\sin\theta$. For this study, we focus on two phenomenologically distinct production regimes:
\begin{itemize}
    \item \textit{Rare $B$-Meson Decays:} For lighter scalars, where the mass of $\phi$, $m_{\phi}$, lies between the $K$-meson and $B$-meson masses, the production of $\phi$ is dominated by the flavor-changing neutral current (FCNC) transition, $b \to s \phi$, which is $\sin^2\theta$ suppressed.
    \item \textit{Exotic Higgs Decays:} For $m_{\phi} < m_h/2$, the SM-like Higgs boson can decay into a pair of dark scalars ($h \to \phi\phi$). An advantage of this production mode is that it depends on the trilinear coupling between the SM-like Higgs boson and $\phi$, which does not need to be suppressed to achieve long lifetimes of $\phi$, as it affects only the branching ratio $\mathrm{Br}(h \to \phi\phi)$.
\end{itemize}
The decay length of $\phi$, $c\tau$, is determined by $m_{\phi}$ and the mixing parameter $\sin\theta$, while its branching ratios to various SM particles depend on $m_\phi$.

\vspace*{-0.4cm}

\subsection{Heavy Neutral Leptons}
\label{ssec:HNL}

\vspace*{-0.3cm}

Heavy neutral leptons, $N$, are right-handed fermionic singlets that mix with the active SM neutrinos through a mixing matrix element $U_{\alpha N}$ ($\alpha = e, \mu, \tau$)\,\cite{Abdullahi:2022jlv}. We specifically benchmark HNL production via the leptonic decays of $D$ and $D_s$ mesons (e.g., $D_s \to \mu N$), where the production rate is governed by $|U_{\mu N}|^2$. Following production, the HNL subsequently decays into SM particles through neutral and charged current interactions. Depending on the mass of $N$, $m_N$, these decays proceed either leptonically or semi-leptonically. 

\section{The \textsf{DELIGHT-SHIELD} detector at a dedicated LLP Interaction Point at FCC-\MakeLowercase{hh}}
\label{sec:dedicatedIP}

Several dedicated detector proposals exist for LLP searches at the FCC-hh, 
such as the transverse detector DELIGHT~\cite{Bhattacherjee:2021rml} and the forward detector FOREHUNT~\cite{Bhattacherjee:2023plj}. 
In this work, we propose a complementary and distinct configuration -- a dedicated IP at the FCC-hh exclusively for LLP searches, equipped with the proposed detector,
\textsf{DELIGHT-SHIELD} (\textbf{De}tector for \textbf{L}ong-l\textbf{i}ved 
particles at hi\textbf{gh} energy of 100 \textbf{T}eV -- 
\textbf{S}hielding \textbf{H}oused \textbf{I}nside \textbf{E}xternal \textbf{L}ayered \textbf{D}etectors).
Similar to the LHC, the FCC-hh collider is proposed to have four different IPs, where two would house general-purpose detectors. We propose that one of these IPs can be dedicated to LLP searches by housing the proposed \textsf{DELIGHT-SHIELD}.
The \textsf{DELIGHT-SHIELD} would consist of a cylindrical composite shielding structure designed to suppress the dominant SM backgrounds, which arise primarily from soft-QCD processes. 
The shielding is proposed to be surrounded by tracking layers. Additional calorimeters can be incorporated to further enhance energy measurements along with the detection of LLP decays to neutral particles.
The composition, dimensions, and attenuation performance of the shielding are discussed in the following sections. 

\vspace*{-0.4cm}

\begin{figure}[h]
\centering
\includegraphics[width=0.65\linewidth]{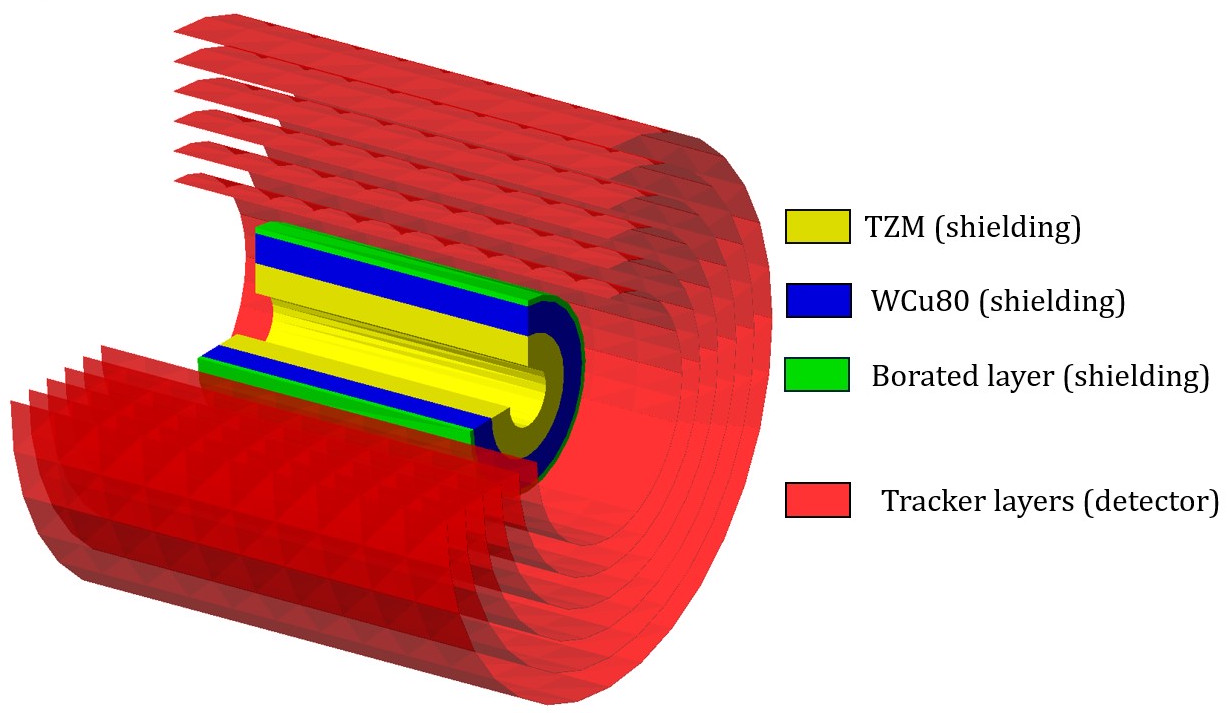}
\caption{Design of the proposed \textsf{DELIGHT-SHIELD} detector.}
\label{det_design} 
\end{figure}

\subsection{Material composition of the shielding for \textsf{DELIGHT-SHIELD}}
\label{ssec:material}

\vspace*{-0.3cm}

For the shielding, we require materials with high melting points to remain structurally stable in the high-multiplicity environment of the FCC-hh with highly energetic particles, and small interaction lengths with the ability to attenuate or absorb most SM electromagnetic and hadronic backgrounds with the minimum possible thickness. To achieve this, we propose a three-layer shielding composed of the following materials.

\begin{itemize}
    \item Motivated by the radiation-hard target design of the SHiP experiment~\cite{lopez2019design}, the first layer is fabricated from \textbf{TZM alloy} (0.5\% titanium, 0.08\% zirconium, and 0.02\% carbon, with the balance being pure molybdenum). TZM is selected for the innermost layer due to its exceptional thermal conductivity and high melting point ($\approx 2623\,^{\circ}$C)\cite{Plansee_TZM}, enabling it to maintain structural integrity under the intense particle flux and heat load adjacent to the IP.

    \item The second layer is designed for maximum stopping power. Two candidate materials are evaluated -- C26000 brass (70\% copper, 30\% zinc; melting range $915$--$955\,^{\circ}$C~\cite{brass_c26000}) and WCu80 composite (80\% tungsten, 20\% copper; melting point $\approx 1083\,^{\circ}$C~\cite{wcu_shielding_2022}). Our simulations indicate that brass produces significantly higher secondary particle yields, as shown in Appendix Table.\ref{tab:brass_W_yield_ratio} -- exceeding those of WCu80 by an order of magnitude in several energy regimes -- which would compromise the zero-background requirement. Consequently, \textbf{WCu80} is selected for its superior electromagnetic and hadronic attenuation properties. It is selected over pure tungsten due to its 
better thermal conductivity, which enhances heat dissipation and reduces thermal gradients within this layer.

    \item The third and final layer consists of a \textbf{boron-loaded polymer composite}, with a density of $1.05\,\text{g/cm}^3$, $85.5\%$ carbon, $9.5\%$ hydrogen, and $5.0\%$ boron. This layer is essential for attenuating the secondary neutron flux generated by hadronic interactions in the preceding metallic stages. The hydrogen content moderates fast neutrons via elastic scattering, while the $5\%$ boron loading ensures efficient thermal neutron capture through the $^{10}$B$(n,\alpha)^7$Li reaction~\cite{marshield_bpe}.
\end{itemize}

The shielding part of the \textsf{DELIGHT-SHIELD} design comprises of three functionally distinct layers -- the TZM, WCu80 and the boronated layer. We will discuss the required thicknesses of these layers to achieve a low-background environment for LLP searches in Section\,\ref{ssec:thickness}.

\vspace*{-0.4cm}

\subsection{The \textsf{DELIGHT-SHIELD} detector}
\label{ssec:delight-shield}

\vspace*{-0.3cm}


The proposed \textsf{DELIGHT-SHIELD} concept at a possible dedicated IP at the FCC-hh consists of an inner shielding system and an outer detector. The materials for the shielding layers were discussed in the previous section, surrounded by an 
outer detector system described here.

Positioned downstream of the shielding, the 
\textsf{DELIGHT-SHIELD} detector consists of multiple concentric 
cylindrical layers optimized for LLP reconstruction 
and residual background rejection. The innermost layers 
are instrumented with silicon pixel detectors, which 
provide the precision vertexing and high-resolution 
tracking required to reject residual SM 
backgrounds and identify displaced decay vertices. 
Crucially, because the composite shield absorbs the 
majority of the hadronic and electromagnetic flux 
before it reaches the tracking volume, radiation 
hardness is a significantly less stringent requirement 
here than in a conventional inner tracker, making 
silicon pixel technology a natural and practical choice 
for this application.

The outer tracking layers are instrumented with 
Resistive Plate Chambers 
(RPCs)\footnote{RPCs typically employ fluorinated gases 
with high global warming potential. Eco-friendly 
alternatives are under active development~\cite{Rigoletti:2023aop} 
and will be prioritised for the \textsf{DELIGHT-SHIELD} design.}, 
which provide a well-established, high-rate capable, 
and cost-effective solution for instrumenting the large 
cylindrical volumes required to reconstruct displaced 
vertices from LLP decays into charged SM particles occurring at significant radial distances 
from the interaction point. The modular nature of the 
RPC system also allows additional layers to be 
incorporated if required by the physics program.

In addition to the tracking layers, calorimeter modules may be interleaved with or appended to the tracking
system to provide complementary energy measurements 
and improve particle identification, further enhancing 
signal-to-background discrimination for LLP decay 
topologies involving photons or neutral hadrons. A 
dedicated timing layer deployed immediately downstream 
of the shield would provide additional discrimination 
power for LLP analyses. Finally, a magnetic 
field, though necessarily modest given the already 
reduced particle momenta downstream of the shield, 
could be incorporated to provide charge identification 
and improved momentum reconstruction, further enhancing the overall sensitivity of the detector.

The detector geometry is 
characterized by three tunable parameters that can be optimized for a 
given physics target -- the inner radius of the cylindrical volume 
($R_{\text{in}}$), the outer radius ($R_{\text{out}}$), and the 
detector length ($L$). The total thickness of the shielding material, placed between the IP and the RPC layers, will determine possible values for the $R_{\text{in}}$ parameter.
The impact of variations in these geometric 
parameters on the sensitivity reach in the LLP parameter space is studied 
systematically in Section\,\ref{ssec:full_lumi}. 
The \textsf{DELIGHT-SHIELD} detector proposed to be installed around a dedicated IP for LLP searches at FCC-hh is illustrated in Fig.~\ref{det_design}.

The combination of large solid angle coverage, proximity 
to the IP, large detection volume, and the significant suppression of electromagnetic and hadronic backgrounds  
provided by the shielding structure enables sensitivity to exceptionally low-rate processes. 
This configuration is expected to yield coverage across a broad region of LLP parameter space, providing critical complementarity to other proposed detectors at the FCC complex.

\vspace*{-0.4cm}

\subsection{Thermal management and heat load estimation}
\label{ssec:thermal}

\vspace*{-0.3cm}

The high particle flux at the IP of FCC-hh leads to heat deposition throughout the shielding volume. To estimate the steady-state operating temperature, we perform a thermal analysis using representative FCC-hh beam parameters.

These calculations are based on soft-QCD processes, which dominate the total energy deposition despite producing lower-energy particles than hard-QCD processes. While hard-QCD interactions generate 
more energetic secondaries, their substantially lower cross-section is expected to result in a subdominant contribution to the integrated heat load compared to the high rate soft-QCD background. The hadron yield is estimated to be about 40 hadrons per inelastic event using \texttt{softQCD} events simulated with \textsc{Pythia\,8}~\cite{Bierlich:2022pfr} at 
$\sqrt{s} = 100\,\text{TeV}$, with the selection criteria 
$|\eta| < 4$ and $p_\text{T} > 0.5\,\text{GeV}$.
Assuming a mean pile-up ($\langle\mu\rangle$) of 1000 $pp$ interactions per bunch crossing and a bunch crossing rate ($f_\text{BC}$) of $40\,\text{MHz}$\,\cite{fcc2019fcc}, the total hadron flux incident on the shield is,
\begin{align}
    \frac{dN}{dt} &= N_\text{hadrons/event} \times 
    \langle\mu\rangle \times f_\text{BC} \nonumber \\
    &= 40 \times 1000 \times 40\,\text{MHz} 
    = 1.6\times10^{12}\,\text{particles\,\,\,sec}^{-1}.
\end{align}
Under the conservative assumption that each hadron carries an average energy of 
$\langle E \rangle = 5\,\text{GeV}$ and is fully absorbed 
by the shield, the total deposited power is,
\begin{align}
    P &= \langle E_\text{hadron} \rangle \times \frac{dN}{dt} 
    \nonumber \\
    &= 5\,\text{GeV} \times 1.6\times10^{12}\,\text{s}^{-1} 
    = 1.28\,\text{kW}.
\end{align}

In the absence of active cooling, and assuming that the shield 
operates in vacuum, heat loss is dominated by thermal 
radiation. At steady-state, the deposited power is balanced 
by radiative emission according to the Stefan-Boltzmann law,
\begin{equation}
    P = \varepsilon\, \sigma\, A\, (T_\text{ss}^4 - T_\text{env}^4),
\end{equation}
where $\varepsilon$ is the surface emissivity, $\sigma = 
5.670 \times 10^{-8}\,\text{W\,m}^{-2}\text{K}^{-4}$ is the 
Stefan-Boltzmann constant, $A$ is the total radiating surface 
area of the shield, $T_\text{ss}$ is the steady-state 
temperature, and $T_\text{env}$ is the ambient environment 
temperature.

For a hollow cylindrical shield with inner radius 
$r_1 = 10\,\text{cm}$, outer radius 
$r_2= 50\,\text{cm}$, and length $l = 2\,\text{m}$, 
the total radiating surface area, comprising the outer 
curved surface and the two annular end caps, is given as,
\begin{align}
    A &= 2\pi r_2 l + 2\pi(r_2^2 - r_1^2) 
    \approx 7.79\,\text{m}^2.
\end{align}
Taking a conservative emissivity of $\varepsilon = 0.1$, $T_\text{env} = 300\,\text{K}$, 
and $P = 1281.6\,\text{W}$, the steady-state condition becomes,
\begin{equation}
    1281.6 = (0.1)\cdot(5.670\times10^{-8})\cdot(7.79)
    \cdot(T_\text{ss}^4 - 300^4).
\end{equation}
Solving for $T_\text{ss}$, we get
\begin{equation}
  T_\text{ss} \approx 439\,\text{K} \quad (\approx 166\,^{\circ}\text{C})
\end{equation}

This operating temperature is well below the melting points of the constituent materials used in the shield (TZM: $2623\,^{\circ}$C, 
WCu80: $1083\,^{\circ}$C), confirming that the thermal load 
from the soft-QCD hadron flux poses no risk of 
structural degradation under purely radiative cooling.

Additionally, nuclear binding energy release from spallation reactions 
may contribute to heat deposition. While titanium and copper have nuclear binding energies of approximately 0.5\,GeV per nucleus, 
molybdenum and zirconium have moderate binding energies of approximately 0.8\,GeV, whereas tungsten exhibits higher binding energy of 1.47\,GeV per nucleus.~\cite{AME2020,KAERI}. 
Under the conservative assumption that each incident hadron completely 
dissociates one tungsten nucleus, the total deposited power increases to 
$P \approx 1.67$\,kW, raising the steady-state temperature to 
$\sim$190\,$^{\circ}$C, still well within acceptable thermal limits for 
the shield materials.

For context, the SHiP fixed-target 
experiment~\cite{lopez2019design} operates with a deposited power of 
305\,kW, which is approximately 183 times higher than our estimate,
necessitating active water cooling to maintain temperatures below 
300\,$^{\circ}$C. While the FCC-hh shield could operate passively, we 
propose incorporating millimeter-scale water-cooling channels at 
2\,cm intervals, following the high-intensity target design strategies 
developed for SHiP. This provides a robust safety margin against 
localized hot spots, long-term activation heat, and thermal cycling 
fatigue, while offering operational flexibility for potential 
higher-luminosity running scenarios.

Another reason for the requirement of active cooling is that the thermal radiation from a 190\,$^{\circ}$C surface would deposit significant 
heat on the downstream silicon pixel detectors, which require operation 
at sub-zero temperatures for optimal performance. Active cooling of the 
shield to $<$50\,$^{\circ}$C is therefore essential not only for thermal 
management of the shield itself, but also to minimize radiative heating 
of adjacent temperature-sensitive detector components.

The dimensional requirements of the shield are discussed in the following section.

\vspace*{-0.4cm}

\subsection{Shielding thickness optimization}
\label{ssec:thickness}

\vspace*{-0.3cm}

Having established the material composition of the shield, we now determine 
the required thickness of each layer. 
The goal of the shielding design is to suppress the dominant soft-QCD backgrounds, produced from the high PU environment, as much as possible by absorbing them within the shield volume. Hard-QCD processes, which produce high-energy, highly penetrating particles, might traverse a shielding structure of realistic thickness with little attenuation and produce even more secondary radiation. They are, therefore, not the target of this shielding strategy. However, the shielding would help in bringing down the energy range of particles produced from the hard-QCD processes as well.

\vspace*{-0.4cm}

\subsubsection*{\textbf{Analytical estimation of suppression with shielding thickness}}
\label{sssec:analytical_thick_shield}

\vspace*{-0.2cm}

Assuming equal-thickness layers of TZM and WCu80, each comprising half of the total shield thickness ($l$), the fraction of incident hadron flux escaping the shield is~\footnote{In this analytic calculation, we neglect the borated polymer layer, as its contribution to the suppression determined by the nuclear interaction length is subdominant.},
\begin{equation}
    F_\text{esc}(l) = \exp\left(-\frac{l}{2}
    \left(\frac{1}{\lambda_\text{TZM}} + 
    \frac{1}{\lambda_\text{WCu80}}\right)\right)
    \label{eq:shield_escape},
\end{equation}
where $\lambda_\text{TZM} = 15.1\,\text{cm}$ and 
$\lambda_\text{WCu80} = 10.9\,\text{cm}$ are the nuclear 
interaction lengths of TZM and WCu80 
respectively~\cite{pdg2024}. This is equivalent to a 
single material shield with an effective interaction 
length of
\begin{equation}
    \lambda_\text{eff} = \frac{2\lambda_\text{TZM}
    \lambda_\text{WCu80}}{\lambda_\text{TZM} + 
    \lambda_\text{WCu80}} \approx 12.66\,\text{cm}.
\end{equation}
The required thickness for various levels of flux 
suppression, obtained by inverting 
Eq.~\eqref{eq:shield_escape}, are presented in 
Table~\ref{tab:shielding_attenuation}.

\begin{table}[htbp]
    \centering
    \caption{Required total shielding thickness as a function of hadronic flux attenuation for a composite TZM–WCu80 shield, assuming equal thicknesses for the two layers.}
    \label{tab:shielding_attenuation}
    \setlength{\tabcolsep}{10pt}
    \begin{tabular}{S[table-format=1.1e-1] S[table-format=3.2]}
        \toprule
        {\% of Flux Escaping} & {Required Thickness [\si{cm}]} \\ 
        \midrule
        1.0e+00 & 58.30 \\
        1.0e-01 & 87.46 \\
        1.0e-02 & 116.61 \\
        1.0e-03 & 145.76 \\
        1.0e-04 & 174.91 \\
        1.0e-05 & 204.07 \\
        \bottomrule
    \end{tabular}
\end{table}

As shown in Table~\ref{tab:shielding_attenuation}, a shield thickness of 
$58.30\,\text{cm}$ attenuates $99\%$ of the incident hadron flux, while 
$116.61\,\text{cm}$ achieves $99.99\%$ attenuation. For a thickness of 
approximately $204\,\text{cm}$, only $10^{-5}\%$ of the original flux 
survives, corresponding to a suppression factor of $10^{-7}$, or 
equivalently $99.99999\%$ attenuation.

From our previous estimate with \textsc{Pythia\,8}, we expect to observe approximately 40 hadrons per soft-QCD event. Scaled by the expected pile-up of 1000 $pp$ interactions per bunch crossing at the FCC-hh~\cite{fcc2019fcc}, this corresponds to approximately 4$\times10^{4}$ hadrons incident on the shield per bunch crossing. A suppression factor of at least $10^{-5}$ is, therefore, mandatory to reduce the incident flux to below one penetrating hadron per bunch crossing. From Table~\ref{tab:shielding_attenuation}, this corresponds to a minimum shield thickness of $145.76\,\text{cm}$, which achieves a suppression of $10^{-5}$ (i.e.\ $99.999\%$ attenuation). 
To ensure a cleaner environment for high-purity signal throughout the full $30\,\text{ab}^{-1}$, we adopt a conservative benchmark depth of $2\,\text{m}$. This achieves a suppression factor of ${\sim}10^{-7}$, effectively reducing the mean hadronic punch-through to well below one particle per bunch crossing, even under peak luminosity conditions.

While these analytical estimates provide a primary baseline, they represent an idealized lower bound as they do not account for secondary particle production within the shield volume. In the following section, we use \textsc{Geant4}\cite{Agostinelli:2002hh,geant4_physlist,Allison:2016lfl} to provide a complete characterization of the shower development, including the production of secondary hadrons, leptons, and photons, to verify the shielding performance in a realistic experimental environment.

\vspace*{-0.4cm}

\subsubsection*{\textbf{Thick shield -- {\rm \textsc{Geant4}} response}}
\label{sssec:thick_shield_geant4}

\vspace*{-0.2cm}

To validate the analytical attenuation estimates of the 
previous section and to characterize 
the secondary particle spectra emerging from the thick 
composite shield, a full \textsc{Geant4} simulation is 
performed. For each primary particle species and energy, $10^5$ 
primary particles are directed into the shield using 
the particle gun utility. The \textsc{Geant4} simulation 
employs the \texttt{FTFP\_BERT} physics list\cite{geant4_physlist}, which 
combines the Fritiof (FTF) string model for high-energy 
hadronic interactions with the Bertini intranuclear 
cascade model for lower energies. All secondary 
particles emerging with energy above $0.5\,\text{GeV}$ 
are recorded. The simulation is performed for three benchmark total shield thicknesses -- 60\,cm, 110\,cm, and 210\,cm. In each case, the shield consists of equal thicknesses of TZM and WCu80 (25\,cm, 50\,cm, and 100\,cm per layer, respectively), followed by a fixed 10\,cm boron-loaded polymer layer for thermal neutron capture. For this study, we assume equal thicknesses of TZM and WCu80; however, this ratio can be further optimized in an actual experimental implementation to enhance particle suppression.
The shield response is evaluated for primary particle energies of 10\,GeV, 20\,GeV, 50\,GeV, 100\,GeV, and 200\,GeV. 
Here, we summarize the total secondary yield per $10^5$ primaries at a representative primary energy of $10\,\text{GeV}$ and $20\,\text{GeV}$ in Table~\ref{tab:thick_shield_split}. These specific energy levels were selected as they characterize the typical maximum energy scales observed in soft-QCD processes.
For completeness, results for individual particle yields after shielding, including higher primary energies of 50\,GeV, 100\,GeV, and 200,GeV, are provided in Tables~\ref{tab:thick_geant4_gamma}--\ref{tab:thick_geant4_mum} in the Appendix. These are included to characterize the shield response over a broader energy range, rather than to reflect the typical soft-QCD particle spectrum.

\begin{table}[htbp]
    \centering
    \caption{Total secondary particle yield per $10^5$ primary particles with $E > 0.5\,\text{GeV}$, for a representative primary energy of $10\,\text{GeV}$ and $20\,\text{GeV}$, as a function of the total shield thickness. The shield thickness values used in the simulation are 60\,cm, 110\,cm, and 210\,cm. The suppression factor in the final column is defined as the ratio of the secondary yield at 60\,cm to that at 210\,cm total depth, quantifying the additional attenuation gained by increasing the shield thickness.}
    \label{tab:thick_shield_split}
    
    \begin{minipage}{0.48\textwidth}
        \centering
        \caption*{(a) $E = 10\,\text{GeV}$}
        \begin{tabular}{l rrr r}
            \toprule
            Primary & \multicolumn{3}{c}{Secondary yield} & Suppression \\
            \cmidrule(lr){2-4}
            particle & 60\,cm & 110\,cm & 210\,cm & (60$\to$210) \\
            \midrule
            $\gamma$   & 15     & 1      & $<1$  & $>10$ \\
            $e^-$      & 6      & $<1$   & $<1$  & $>10$ \\
            $\pi^{\pm}$& 31,814 & 1,966  & 290   & $\sim$110 \\
            $p$        & 34,428 & 1,823  & 17    & $\sim$2000 \\
            $n$        & 36,203 & 1,916  & 9     & $>1000$ \\
            $K^{\pm}$  & 37,287 & 3,095  & 962   & $\sim$39 \\
            $\mu^-$    & 100,224 & 100,154 & 100,007 & $\sim$1 \\
            \bottomrule
        \end{tabular}
    \end{minipage}
    \hfill
    \begin{minipage}{0.48\textwidth}
        \centering
        \caption*{(b) $E = 20\,\text{GeV}$}
        \begin{tabular}{l rrr r}
            \toprule
            Primary & \multicolumn{3}{c}{Secondary yield} & Suppression \\
            \cmidrule(lr){2-4}
            particle & 60\,cm & 110\,cm & 210\,cm & (60$\to$210) \\
            \midrule
            $\gamma$   & 39     & 4      & $<1$  & $>10$ \\
            $e^-$      & 15     & $<1$   & $<1$  & $>10$ \\
            $\pi^{\pm}$& 78,904 & 5,865  & 238   & $\sim$330 \\
            $p$        & 80,878 & 5,118  & 90    & $\sim$900 \\
            $n$        & 83,031 & 5,463  & 70    & $\sim$1200 \\
            $K^{\pm}$  & 84,097 & 5,875  & 680   & $\sim$124 \\
            $\mu^-$    & 100,412 & 100,419 & 100,352 & $\sim$1 \\
            \bottomrule
        \end{tabular}
    \end{minipage}
\end{table}


Three primary conclusions emerge from the simulation. 
First, electromagnetic primaries ($\gamma$, $e^-$) are 
effectively eliminated within $60\,\text{cm}$, 
consistent with the radiation-length expectations for high-$Z$ materials. 
Second, hadronic primaries ($\pi^{\pm}$, $p$, $n$, $K^{\pm}$) undergo significant shower multiplication at shallow depths and hence don't show much suppression. However, at the benchmark depth of $210\,\text{cm}$, the secondary 
yield is suppressed by a factor of 
$\mathcal{O}(40$--$2000)$ and $\mathcal{O}(120$--$1200)$ relative to $60\,\text{cm}$ for $10\,\text{GeV}$ and $20\,\text{GeV}$ primary energies, respectively. 
Third, muons are essentially unattenuated at all values of shield thicknesses. 
This confirms that while the shield is transparent to muons, it provides a robust hadronic and electromagnetic suppression required to maintain a low-occupancy environment in the downstream tracking layers.

The results presented above characterize the general 
shield response for mono-energetic primaries of a given 
species, each simulated with $10^5$ particles of equal 
weight. In practice, however, the actual particle 
multiplicities and species composition incident on the 
shield are those of the soft-QCD flux, in which 
different particle species are present in very different 
relative abundances and with a steeply falling energy 
spectrum. To obtain a realistic estimate of the residual background, the simulation is performed for soft-QCD 
processes at $\sqrt{s} = 100\,\text{TeV}$ with kinematic 
selection criteria $|\eta| < 4$, $p_\text{T} > 0.5\,\text{GeV}$, 
and a mean pile-up of $\langle\mu\rangle =$1000.
The incident particle spectra are first generated using 
\textsc{Pythia\,8} with the above 
kinematic selections. The energy spectrum of each particle species is binned in $1\,\text{GeV}$ intervals, with bin centers ranging from $1$ to $20\ \text{GeV}$.The dominant particle 
contributions to the soft-QCD flux are tabulated in Table~\ref{tab:softqcd_results_100TeV} in the Appendix. As expected from the steeply falling $p_\text{T}$ spectrum of soft-QCD processes, the lowest energy bin carries the largest particle multiplicity, with yields falling rapidly at higher energies.
For each particle species and energy bin in the table, the \textsc{Geant4} shield response with the thickness of 60\,cm, is simulated using the 
particle gun utility, with particles fired at the central energy of each bin. The number of secondary particles emerging from the shield with energy above $0.1\,\text{GeV}$ is recorded and scaled by the corresponding primary particle multiplicity from the \textsc{Pythia\,8} soft-QCD simulation. 
The results are presented in 
Table~\ref{tab:50cm_thick_shield_reponse_100TeV} in the Appendix.

It should be noted that while the final total secondary 
particle yields presented in 
Table~\ref{tab:thick_shield_split} or energy-binned weighted with the corresponding soft-QCD primary yields in Table\,\ref{tab:50cm_thick_shield_reponse_100TeV} of the Appendix are not zero, 
the energy distribution of these secondaries falls steeply with energy. The majority of secondary particles carry energies of only a few GeV, 
reflecting the soft hadronic shower spectrum generated 
within the shield. 
Consequently, while the residual flux downstream of 
the shield cannot be reduced to zero, the application 
of a modest energy threshold on the downstream detector, 
combined with the precision background rejection 
capabilities of the silicon pixel tracking layers is expected to bring it well within manageable limits for LLP analyses.


Comparing the total pion flux incident on the 60\,cm shield 
with that emerging downstream, a suppression of 
approximately $98\%$ is observed, on comparing the initial soft-QCD fluxes in Table\,\ref{tab:softqcd_results_100TeV} with the particle yield after the shielding in Table\,\ref{tab:50cm_thick_shield_reponse_100TeV}, consistent with the 
analytical attenuation estimates of 
Eq.~\eqref{eq:shield_escape} and 
Table~\ref{tab:shielding_attenuation}. Similar 
suppression factors are observed for charged and neutral kaons.
For nucleons, the secondary proton and neutron yields 
downstream of the shield are somewhat higher than 
analytically expected. This excess arises from hadronic 
shower development within the shield, in which 
inelastic interactions of the primary hadrons produce 
additional low-energy nucleons as spallation products. 
However, these secondary nucleons are predominantly 
very low energy and are therefore strongly suppressed 
by the application of an energy threshold on the 
downstream detector. This is illustrated by the neutron 
flux as a function of energy threshold -- after a 60\,cm shield, at a threshold 
of $0.1\,\text{GeV}$, only $5\%$ of the neutron flux 
is suppressed as compared to the primary flux; raising the threshold to $0.5\,\text{GeV}$ 
increases the suppression to $90\%$; at $1\,\text{GeV}$ 
it reaches $98\%$; and at $2\,\text{GeV}$ more than 
$99\%$ of the primary neutron flux is eliminated. 
This demonstrates that while the shield alone cannot 
reduce the SM backgrounds to zero, a modest energy 
threshold applied downstream is sufficient to render 
the residual background negligible for LLP searches.
To quantify this concretely, the total soft-QCD hadron 
flux incident on the shield is approximately $4 \times 
10^4$ hadrons per bunch crossing, for selection criteria 
$|\eta| < 4$ and $p_\text{T} > 0.5\,\text{GeV}$ at a 
pile-up of $\langle\mu\rangle = 1000$, as tabulated in 
Appendix Table~\ref{tab:softqcd_results_100TeV}. After 
traversing the $60\,\text{cm}$ composite shield, this 
is reduced to approximately 660 hadrons per bunch 
crossing with $E > 0.5\,\text{GeV}$, corresponding to 
a suppression factor of $\sim$60.




\vspace*{-0.4cm}

\subsection{Sensitivity to the LLP benchmarks}
\label{ssec:full_lumi}

\vspace*{-0.3cm}

In this section, we present projected sensitivity limits for the dark scalar and HNL models, comparing the \textsf{DELIGHT-SHIELD} performance against the FCC-hh barrel muon spectrometer (MS) baseline.
The FCC-hh barrel muon spectrometer, as specified in the reference 
detector design~\cite{fcc2019fcc}, has a half-length of $10\,\text{m}$ 
along the beam axis, an inner radius of $R_\text{in} = 7\,\text{m}$, 
and an outer radius of $R_\text{out} = 8.5\,\text{m}$. For the \textsf{DELIGHT-SHIELD} detector, the half-length is fixed at $10\,\text{m}$, while the inner and outer radii are varied across several configurations to study the dependence of sensitivity on detector geometry. The outer radius $R_\text{out}$ is 
constrained in practice by the dimensions of the 
cavern housing the detector; in this study we extend it 
up to $R_\text{out} = 20\,\text{m}$.  In the previous section, we have found out that even the 60\,cm shield is quite effective in suppressing the SM backgrounds, however, we select the minimum thickness of the shield to be around 1.5\,m as a conservative measure.
By utilizing a 1.5\,m composite shield to suppress the dominant SM backgrounds, 
we evaluate a minimum inner radius of $R_\text{in} = 1.5\,\text{m}$, bringing the decay volume significantly closer to the IP than the MS. For scenarios requiring further background suppression through additional 
shielding, we also benchmark a configurations with $R_\text{in} = 4\,\text{m}$.


We simulate the LLP events based on the three benchmark production modes in $pp$ collisions at $\sqrt{s} = 100\,\text{TeV}$ using \textsc{Pythia\,8}~\cite{Bierlich:2022pfr}.
For the dark scalar production from Higgs boson decay, we consider the dominant gluon fusion production mode for the SM-like Higgs boson. 
In cases where the LLP is produced from meson decays for the dark scalar and the HNL, we produce the meson flux from soft-QCD simulated at $\sqrt{s} = 100\,\text{TeV}$ with \textsc{Pythia\,8}. Subsequently, SM Higgs bosons or mesons generated in \textsc{Pythia\,8} are decayed into $\phi$ according to the specified decay channels in order to obtain the $\phi$ spectrum.
For each parameter space point, we compute the fraction of events in which at least 
one of the LLP decays within the fiducial volume of 
the \textsf{DELIGHT-SHIELD}, constituting a signal event. 
The sensitivity contour is then defined by the requirement of three signal events at an integrated luminosity of $30\,\text{ab}^{-1}$~\cite{fcc2019fcc}, assuming the shielding achieves a near-zero background environment.

\begin{figure*}[hbt!]
\centering
\includegraphics[width=0.49\linewidth]{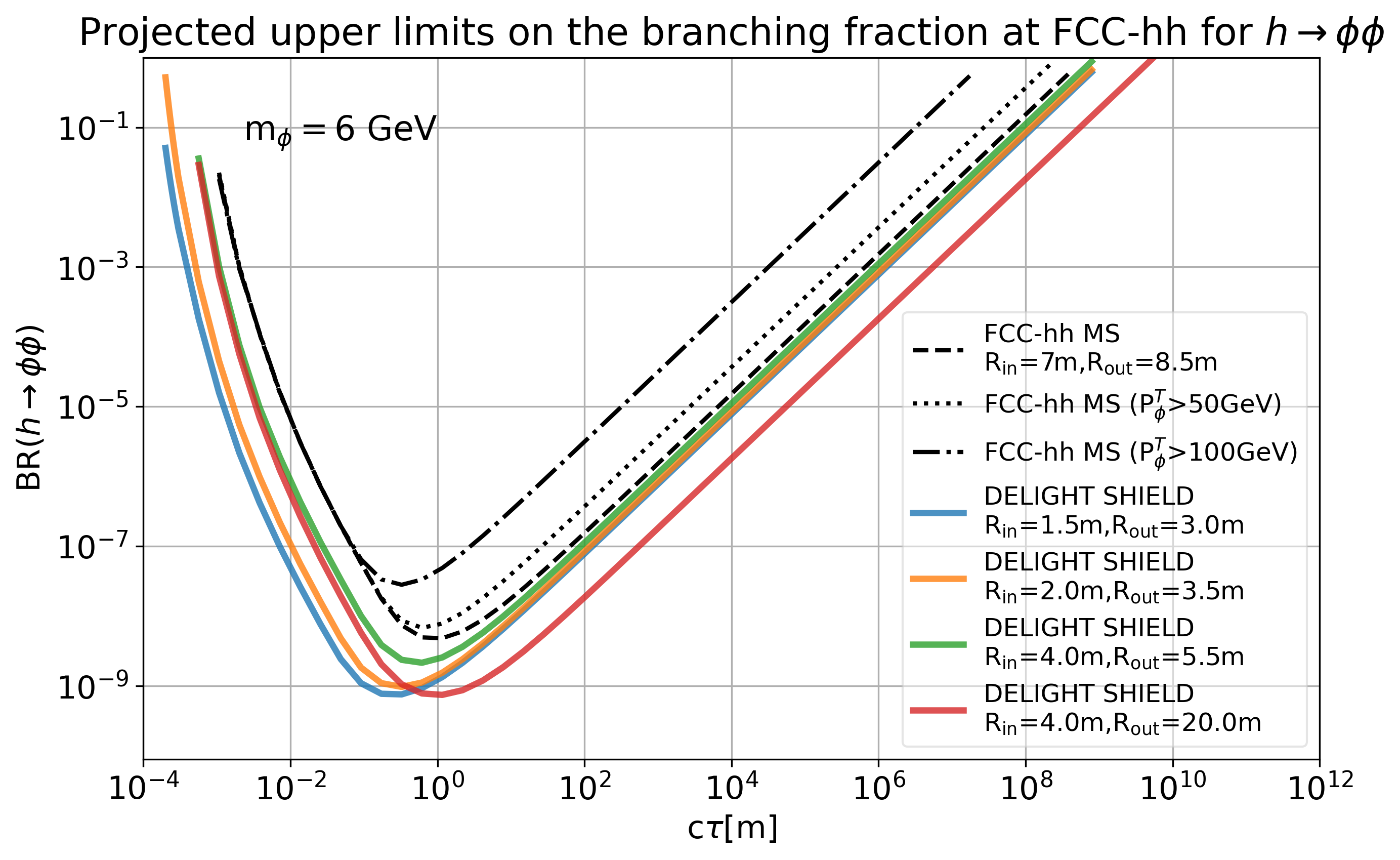}~
\includegraphics[width=0.49\linewidth]{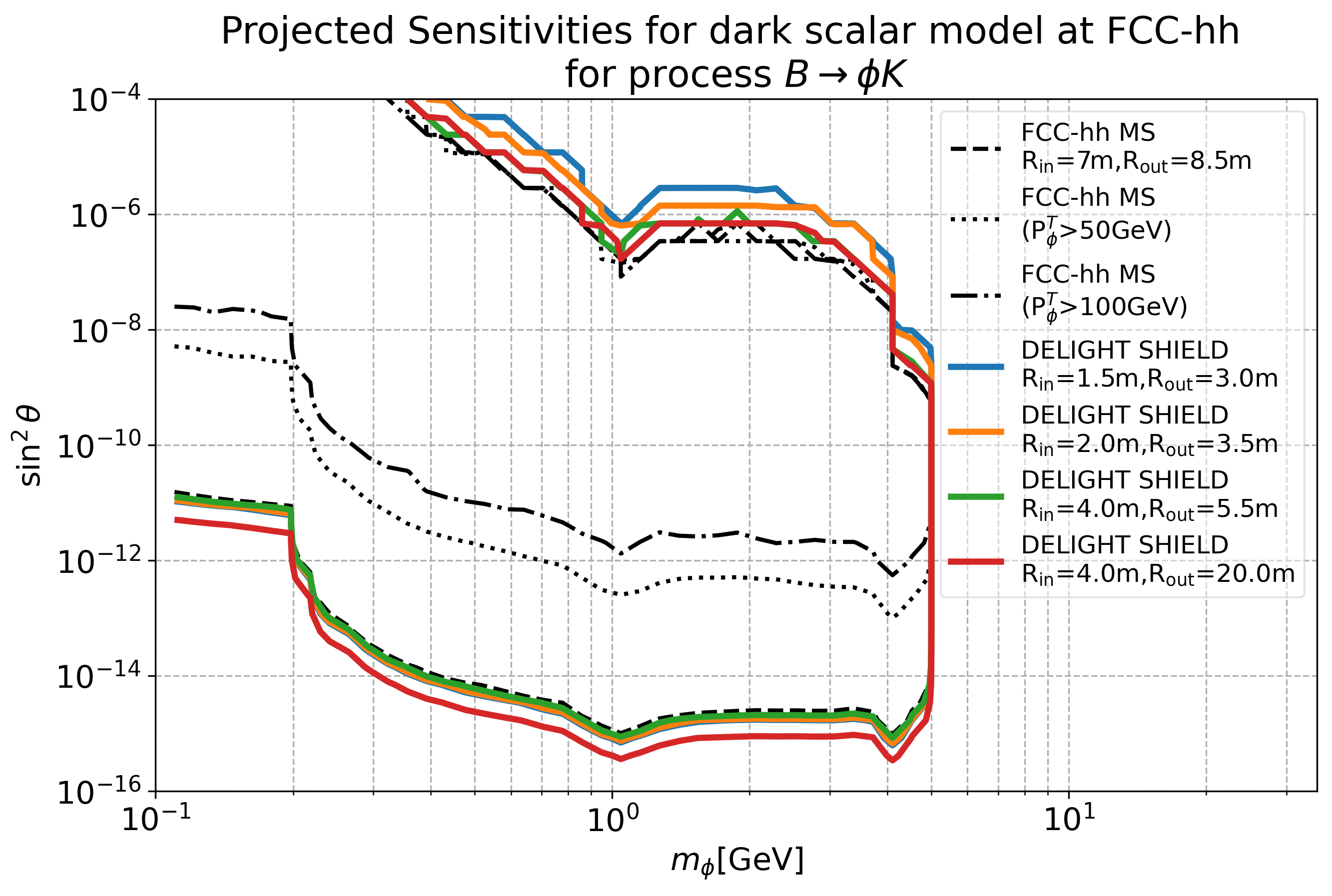}\\
\includegraphics[width=0.49\linewidth]{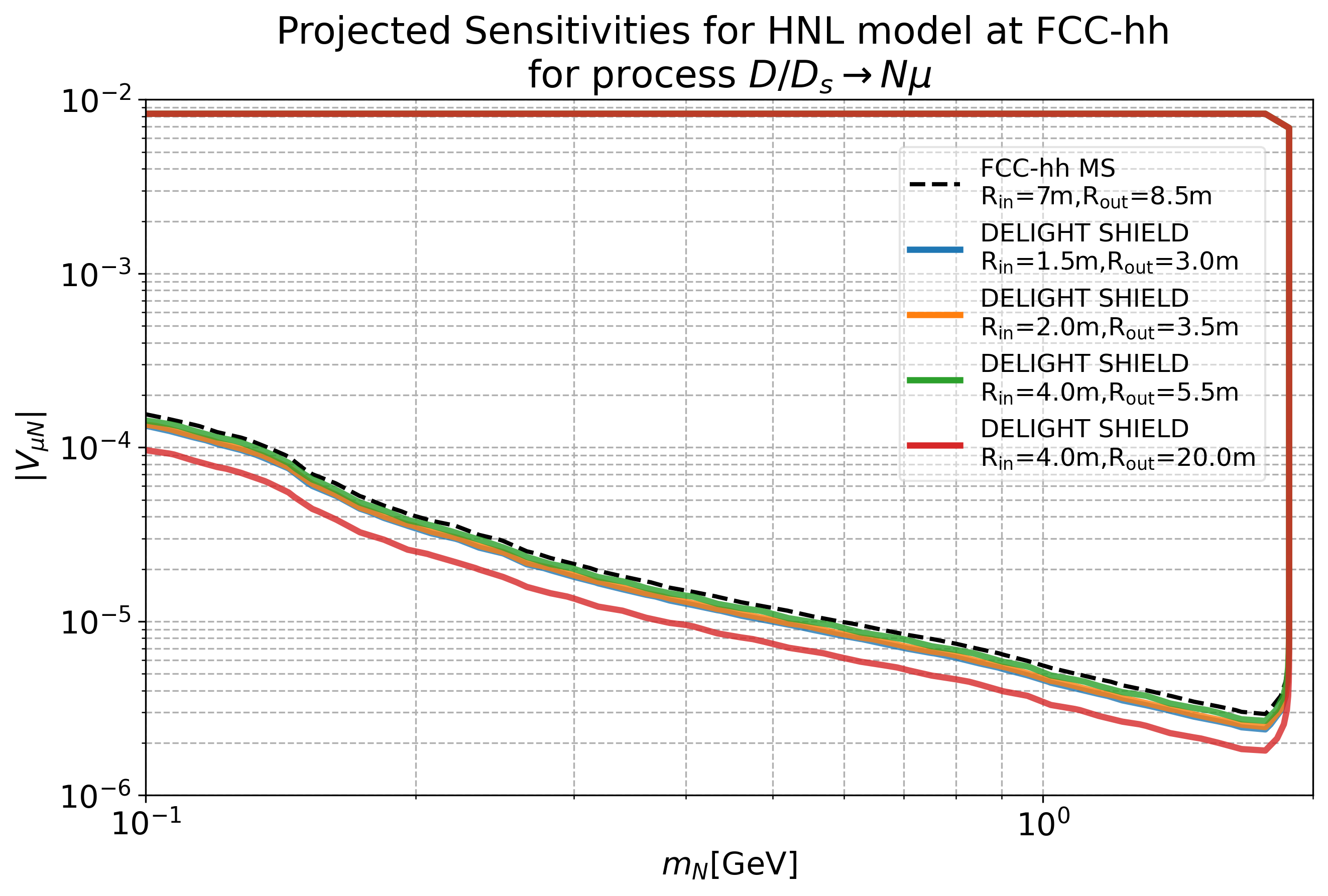}
\caption{Projected sensitivities at the FCC-hh 
($\sqrt{s} = 100\,\text{TeV}$, $\langle\mu\rangle = 1000$) with an integrated 
luminosity of $30\,\text{ab}^{-1}$ 
for three benchmark processes in the dark scalar and HNL 
models, shown for the FCC-hh barrel muon spectrometer and 
for several geometric configurations of the 
\textsf{DELIGHT-SHIELD} detector. 
\textbf{(a)} {\it top left}:~Projected sensitivity in the 
$(c\tau,\,\text{BR}(h \rightarrow \phi\phi))$ plane 
for the process $h \rightarrow \phi\phi$, with the dark 
scalar mass fixed at $m_\phi = 6\,\text{GeV}$. 
\textbf{(b)} {\it top right}:~Projected sensitivity in the 
$(\sin^2\theta,\, m_{\phi})$ plane for the process 
$B \rightarrow \phi K$. 
In both (a) and (b), muon spectrometer sensitivity curves 
are shown for three representative $p_\text{T}^{\phi}$ 
thresholds of 0, 50, and $100\,\text{GeV}$ applied to 
the dark scalar, while no $p_\text{T}$ requirement is 
applied to \textsf{DELIGHT-SHIELD}. 
\textbf{(c)} {\it bottom}:~Projected sensitivity in the 
$(|V_{\mu N}|^2,\, m_{N})$ plane for the process 
$D/D_s \rightarrow N\mu$ in the HNL model. No 
$p_\text{T}$ threshold is applied to either the muon spectrometer or \textsf{DELIGHT-SHIELD}, as the soft $p_\text{T}$ spectrum of the $D/D_s$ parent meson renders high $p_\text{T}$ thresholds ineffective for this production channel.}
\label{h2phiphi_sensitivity}
\end{figure*}


Triggering on LLP signatures in the LHC muon spectrometer traditionally requires 
moderate to high transverse momentum ($p_\text{T}$) 
thresholds to suppress SM backgrounds and manage high trigger rates \cite{mitridate2023energetic}. Extrapolating to the FCC-hh, it is reasonable to expect that similar threshold requirements will be necessary.
While the precise thresholds remain under study, we characterize
the sensitivity loss 
by evaluating three scenarios for the muon spectrometer -- a baseline with no $p_\text{T}$ requirement, and cases with $p_\text{T} > 50\,\text{GeV}$ and $p_\text{T} > 100\,\text{GeV}$ 
applied to the LLP. For the benchmark processes $h \rightarrow \phi\phi$ and $B \rightarrow \phi K$, sensitivity curves for the FCC-hh muon spectrometer are shown at three $p_\text{T}^{\phi}$ thresholds of 0, 50, and $100\,\text{GeV}$. For the HNL model, only the muon spectrometer sensitivity with no $p_\text{T}$ requirement is presented, as imposing thresholds of 50 or $100\,\text{GeV}$ yields no meaningful sensitivity for this channel. This is a consequence of the kinematics of HNL production -- the HNL is produced via two-body leptonic decays of $D$ and $D_s$ mesons, 
which are themselves produced in soft-QCD interactions with characteristically low transverse momentum. As shown in Ref.~\cite{BHATTACHERJEE2026140121} (Appendix Fig.~5), the $p_\text{T}$ distribution of $D$ mesons from soft-QCD is significantly softer than that of $B$ mesons, so the HNL decay products inherit a similarly soft $p_\text{T}$ spectrum and fail to pass high threshold requirements. 

In contrast, the \textsf{DELIGHT-SHIELD} detector is designed to operate in a low-background environment, significantly reducing or eliminating the need for high-$p_\text{T}$ trigger requirements. By utilizing precision tracking and timing in a shielded volume, the experiment maximizes the accessible LLP parameter space, particularly for lower-mass particles or those with soft decay products.
This further illustrates the advantage of \textsf{DELIGHT-SHIELD}, which requires low kinematic threshold and therefore retains full sensitivity to LLP models with soft production kinematics.

The resulting sensitivity contours in the $(c\tau,\,\text{BR})$ plane 
for the FCC-hh barrel muon spectrometer and for several geometric 
configurations of the \textsf{DELIGHT-SHIELD} detector are shown in the {\it top left} panel of Fig.\,\ref{h2phiphi_sensitivity}. For certain configurations, the 
\textsf{DELIGHT-SHIELD} detector achieves sensitivity to branching ratios as 
low as $\mathcal{O}(10^{-9})$ at $c\tau \approx 1\,\text{m}$, 
representing a significant improvement over the muon spectrometer alone.
Analogous sensitivity projections for the $B \rightarrow \phi K$ 
process in the dark scalar model and for the $D/D_s \rightarrow \mu N$ 
process in the HNL model are also presented in the {\it top right} and {\it bottom} panels of Fig.\,\ref{h2phiphi_sensitivity}, 
respectively. 
The sensitivity gain, defined as the ratio of the \textsf{DELIGHT-SHIELD} reach to that of the FCC-hh Barrel muon spectrometer, is shown in Fig.~\ref{comparisons} in the Appendix for the dark scalar produced from Higgs boson decay, evaluated for two \textsf{DELIGHT-SHIELD} detector configurations.
Collectively, these results demonstrate that a dedicated LLP detector such as \textsf{DELIGHT-SHIELD} can provide substantially improved sensitivity, relative to the FCC-hh muon spectrometer, across diverse LLP production modes and models.

\subsubsection*{\textbf{Why \textsf{DELIGHT-SHIELD} outperforms the FCC-hh barrel muon spectrometer?}}
\label{subsubsec:DELIGHT_better_than_MS}

The sensitivity projections presented in Section~\ref{ssec:full_lumi} are necessarily idealized, as they are based purely on geometric decay probability within the fiducial detector volume and do not account for the full complexity of a real experimental environment. 
In practice, additional factors -- such 
as trigger requirements, reconstruction efficiencies, 
and the need to reject residual SM 
backgrounds -- will further affect the \textsf{DELIGHT-SHIELD} and the MS performances.
Several of these factors are expected to enhance the advantage of \textsf{DELIGHT-SHIELD} over the muon spectrometer beyond what is already indicated by the geometric projections alone.
In particular, the MS may require 
high $p_\text{T}$ thresholds on the LLP signal to suppress hadronic punch-through backgrounds -- the sensitivity loss associated with such requirements is already illustrated in Section~\ref{ssec:full_lumi} -- whereas \textsf{DELIGHT-SHIELD}, by virtue of its composite shielding, 
can operate with minimal kinematic requirements. Here we discuss the three principal reasons why \textsf{DELIGHT-SHIELD} offers an improved sensitivity to LLP searches compared to the FCC-hh barrel muon spectrometer.

\textbf{Geometric acceptance and lifetime coverage:} 
The \textsf{DELIGHT-SHIELD} is significantly closer to 
the interaction point than the FCC-hh MS, which has 
an inner radius of $R_\text{in} = 7\,\text{m}$. This 
proximity provides enhanced sensitivity to LLPs with 
shorter proper lifetimes, which predominantly decay 
before reaching the muon spectrometer. Furthermore, 
the outer radius $R_\text{out}$ of \textsf{DELIGHT-SHIELD} can 
be freely extended to increase the volume of fiducial decay, providing a greater geometric acceptance than 
the MS for LLPs with longer lifetimes. The combined 
effect is a broader coverage of LLP lifetime parameter 
space, as demonstrated by the sensitivity projections 
of Section~\ref{ssec:full_lumi}.

\textbf{Hadronic background suppression:} The combined 
nuclear interaction length presented by the ECAL and 
HCAL to particles reaching the FCC-hh muon spectrometer 
is approximately $10.5\,\lambda_I$~\cite{fcc2019fcc}. 
The composite shield of \textsf{DELIGHT-SHIELD}, by contrast, 
has an effective nuclear interaction length of 
$\lambda_\text{eff} = 12.66\,\text{cm}$ 
(Section~\ref{ssec:thickness}), so that the benchmark 
$2\,\text{m}$ thick shielding corresponds to 
$200\,\text{cm}/12.66\,\text{cm} \approx 15.8\,\lambda_I$. 
The additional $5.3\,\lambda_I$ of hadronic absorption 
relative to the FCC-hh main detector translates to an 
enhancement in hadronic suppression of 
$e^{5.3} \approx 200$, meaning the composite shield 
suppresses hadronic backgrounds 
approximately 200 times more effectively than the 
FCC-hh calorimeter system does for the muon spectrometer.

\textbf{Downstream tracking quality:} The FCC-hh muon 
spectrometer is instrumented primarily with RPC-based 
tracking layers. In \textsf{DELIGHT-SHIELD}, by contrast, the 
first few tracking layers immediately downstream of the 
shield are instrumented with silicon pixel detectors 
(Section~\ref{ssec:delight-shield}), which provide 
substantially superior spatial resolution and vertexing 
precision. This enables more effective rejection of 
residual backgrounds that survive the shield, and 
improves the reconstruction quality of displaced 
vertices from LLP decays.

The combination of broader geometric acceptance, 
enhanced sensitivity to shorter LLP lifetimes, 
approximately 200 times greater hadronic suppression, 
and superior downstream tracking resolution makes 
\textsf{DELIGHT-SHIELD} a significantly more capable instrument 
for LLP searches than the FCC-hh barrel muon 
spectrometer alone. It is worth emphasising that the 
second and third points are closely connected -- the 
greater hadronic suppression provided by the composite 
shield, combined with the improved background rejection 
of the silicon pixel tracker, together reduce the 
residual SM background to a level at which 
\textsf{DELIGHT-SHIELD} can operate with substantially lower 
kinematic thresholds on the LLP signal than are 
required by the muon spectrometer. Since the muon 
spectrometer must rely on the FCC-hh calorimeter 
system alone for background suppression, it may requires  $p_\text{T}$ thresholds on the LLP to 
remain background-free, at the cost of significant 
sensitivity loss in the low-$p_\text{T}$ regime. 
\textsf{DELIGHT-SHIELD} is designed to eliminate precisely 
this trade-off, maximizing the accessible LLP 
parameter space by combining physical background 
suppression with precision downstream reconstruction.
\section{Alternative scenarios without a dedicated LLP IP at the FCC-\MakeLowercase{hh}}
\label{sec:no_dedicatedIP}

Our proposed \textsf{DELIGHT-SHIELD} detector, designed for LLP searches at a dedicated IP, assumes that one of the four FCC-hh IPs can be allocated for this purpose throughout the run. The shielding-based detection concept can also be adapted for alternative configurations if a dedicated LLP IP is not available.

\vspace*{-0.4cm}

\subsection{Reduced luminosity runs with \textsf{DELIGHT-SHIELD}}
\label{ssec:reduced_lumi}

\vspace*{-0.3cm}

The preceding analysis established the sensitivity reach of \textsf{DELIGHT-SHIELD} assuming a full FCC-hh integrated luminosity of $30\,\text{ab}^{-1}$~\cite{fcc2019fcc}. 
While this scenario provides the maximum discovery potential, we now investigate the operational impact of a reduced luminosity run. 
This addresses scenarios where the \textsf{DELIGHT-SHIELD} operates for only a portion of the FCC-hh program at an IP, before being transitioned to a general-purpose detector.

In the zero-background regime assumed throughout this analysis, 
the minimum detectable branching ratio for the $h\rightarrow\phi\phi$ process, scales as the inverse of 
the integrated luminosity, $\text{BR}_\text{min} \propto 
\mathcal{L}^{-1}$. However, the superior geometric acceptance and lower $p_\text{T}$
thresholds of \textsf{DELIGHT-SHIELD} provide a luminosity-equivalent gain that can compensate for shorter run times. Even with a fraction of the total luminosity, the shielded design can match or surpass the sensitivity of a general-purpose muon spectrometer operating at full capacity.

Table~\ref{tab:diff_configs_gain} quantifies these results for 
the benchmark process $h \rightarrow \phi\phi$. For each geometric 
configuration of \textsf{DELIGHT-SHIELD}, we report the gain in sensitivity obtained using the \textsf{DELIGHT-SHIELD} detector for the full FCC-hh run along with the integrated luminosity at which \textsf{DELIGHT-SHIELD} achieves sensitivity 
equivalent to the FCC-hh muon Barrel spectrometer LLP search with the full-luminosity ($30\,\text{ab}^{-1}$), evaluated at three representative 
$p_\text{T}$ thresholds of 0, 50, and $100\,\text{GeV}$ 
applied to the LLP signal in the muon spectrometer. Sensitivity gain of \textsf{DELIGHT-SHIELD} over FCC-hh MS, as a function of LLP lifetime are shown in Appendix Fig.\ref{comparisons} for two geometric configurations.
For the large-radius configuration ($R_\text{in} = 4\,\text{m}$, 
$R_\text{out} = 20\,\text{m}$), integrated luminosities of 
$3.57$, $1.48$, and $0.17\,\text{ab}^{-1}$ are sufficient to 
match the muon spectrometer sensitivity at $p_\text{T}$ 
thresholds of 0, 50, and $100\,\text{GeV}$, respectively. For 
the compact configuration ($R_\text{in} = 1.5\,\text{m}$, $R_\text{out} = 3.0\,\text{m}$), the corresponding luminosities 
are $15.35$, $6.35$, and $0.75\,\text{ab}^{-1}$, respectively. 

Remarkably, in the scenario where the muon spectrometer is limited by a $p_\text{T} =100\,\text{GeV}$ trigger, the large-radius \textsf{DELIGHT-SHIELD} requires only $0.17\,\text{ab}^{-1}$ -- representing less than $0.6\%$ of the total FCC-hh runtime -- to reach the same physics goals. Even without a $p_{\rm T}$ cut on the MS based LLP search, the \textsf{DELIGHT-SHIELD} achieves a sensitivity gain greater than unity over the former, owing to its improved geometric acceptance.
These results demonstrate that a modest, dedicated LLP run is not only a viable alternative but a highly efficient use of beam time, offering discovery potential that is robust against the realistic trigger constraints of searches with general-purpose detectors.

\begin{table}[ht]
    \centering
    \caption{Sensitivity gain of \textsf{DELIGHT-SHIELD} over the FCC-hh barrel MS and the integrated luminosity required for \textsf{DELIGHT-SHIELD} to match the 
    full-luminosity ($30\,\text{ab}^{-1}$) MS sensitivity, 
    for several geometric configurations and three 
    $p_\text{T}$ thresholds applied to the LLP signal in 
    the MS. Results are shown for the benchmark process 
    $h \rightarrow \phi\phi$ with $m_\phi = 6\,\text{GeV}$.}
    \label{tab:diff_configs_gain}
    \setlength{\tabcolsep}{12pt}
    \begin{tabular}{cc rrr rrr}
        \toprule
        $R_{\text{in}}$ & 
        $R_{\text{out}}$ & 
        \multicolumn{3}{c}{Sensitivity gain} & 
        \multicolumn{3}{c}{Required luminosity [ab$^{-1}$]} \\
        \cmidrule(lr){3-5} \cmidrule(lr){6-8}
        [m] & [m] & $p_{T,\text{LLP}}^{\rm MS}>$ 0\,GeV & 50\,GeV & 100\,GeV & 
            $p_{T,\text{LLP}}^{\rm MS}>$ 0\,GeV & 50\,GeV & 100\,GeV \\
        \midrule
        1.5 & 3.0 & 1.95 & 4.72 & 40.21 & 15.35 & 6.35 & 0.75 \\
        2 & 3.5  & 1.80 & 4.35  & 37.04  & 16.66 & 6.89 & 0.81 \\
        4 & 5.5  & 1.37 & 3.32  & 28.27  & 21.83 & 9.03 & 1.06 \\
        4 & 20   & 8.41 & 20.32 & 172.94 & 3.57  & 1.48 & 0.17 \\
        \bottomrule
    \end{tabular}
\end{table}

\vspace*{-0.4cm}

\subsection{Shielding to understand anomalous activities in the muon spectrometer}
\label{ssec:improved_MS_analysis_FCC}

\vspace*{-0.3cm}

Consider the scenario where neither a dedicated LLP IP nor a reduced-luminosity run with the \textsf{DELIGHT-SHIELD} is approved at the FCC-hh. A complementary approach to LLP searches can nonetheless be pursued using the shielding design described above. Suppose the FCC-hh MS records anomalous hits 
-- a signature consistent with either hadronic jet punch-through, which constitutes an irreducible SM background, or genuine LLP decays occurring downstream of the inner detector. Distinguishing between these two hypotheses is non-trivial in the standard detector configuration.

A clean experimental test can be constructed as follows. 
If the tracker surrounding the interaction point were 
replaced by approximately $50\,\text{cm}$ of the 
composite shielding material described above, the 
incident soft-QCD hadron flux would be suppressed by 
${\sim}98\%$ according to the analytical estimates of 
Table~\ref{tab:shielding_attenuation}. It should be 
noted, however, that this represents the analytically 
expected suppression for the soft-QCD energy regime, 
and the actual suppression achieved for punch-through 
jets will differ for several reasons. First, punch-through 
jets are by definition high-energy objects, and while 
the nuclear interaction length has only a weak energy 
dependence, the hadronic shower profile and secondary 
particle multiplicity both evolve with energy, as 
illustrated in Appendix 
Tables~\ref{tab:thick_geant4_pip}--\ref{tab:thick_geant4_kaonp}. 
Second, the inelastic interactions of high-energy 
hadrons within the shield produce a significant 
secondary particle flux, which partially repopulates 
the downstream particle spectrum. The actual suppression 
factor for punch-through jets is therefore 
energy-dependent and will in general be lower than the 
soft-QCD estimate. 

Nevertheless, whatever the actual suppression achieved -- which can be further enhanced by imposing higher energy thresholds in the downstream MS -- it will result in reduction of the SM punch-through background, as compared to the one without the shielding material.
While LLPs, which interact only weakly with the shield, would 
pass through largely unaffected. This configuration enables a powerful null-hypothesis test -- 
\begin{itemize}
\item If the rate of anomalous MS hits decreases proportionally with the suppression factor 
proportional to that expected in SM, the observed excess is statistically consistent with SM hadronic punch-through.
\item Conversely, if the signal rate remains approximately invariant under the addition of shielding, the events are incompatible with hadronic backgrounds, providing a robust discovery handle for BSM physics.
\end{itemize}
This ``shield-in/shield-out" methodology effectively transforms the physical shielding into a diagnostic discriminator, allowing for high-confidence LLP searches within the existing FCC-hh HCAL and MS subsystems without requiring specialized run conditions.

\vspace*{-0.4cm}

\subsection{A thin shielding for ECAL background reduction}
\label{ssec:thin_shielding_FCC}

\vspace*{-0.3cm}

In scenarios where a thick shielding (50--200\,cm) is not 
feasible, a thin (${\sim}10\,\text{cm}$) 
composite shield can still provide a significant discovery advantage by suppressing
the soft-QCD photon background, which 
constitutes a substantial source of noise in the ECAL-based searches for LLPs, 
given the large pile-up at FCC-hh. This scenario does not require removing the entire central tracker. 
Only the pixel detector extending to $r = 20$\,cm~\cite{fcc2019fcc} 
needs to be removed, either completely or partially, while the outer 
tracking layers remain intact.

Since the ratio of nuclear interaction length to radiation length 
($\lambda_\text{int}/X_0$) is typically 20--30 for high-$Z$ 
materials~\cite{pdg2024}, a shield of only ${\sim}10\,\text{cm}$ 
is far too thin to attenuate the hadronic component of the 
soft-QCD flux -- primarily charged pions ($\pi^{\pm}$) -- 
to any appreciable degree. However, the electromagnetic component, 
predominantly photons, can be suppressed efficiently over this 
depth. The photon survival fraction as a function of shield 
thickness $t$ (expressed in units of radiation length $X_0$) 
follows from the pair-production mean free path, shown in Table\,\ref{tab:photon_shielding_attenuation}.

\begin{table}[htbp]
    \centering
    \caption{Photon survival fraction as a function of shield 
             depth in units of radiation length $X_0$.}
    \label{tab:photon_shielding_attenuation}
    \begin{tabular}{S[table-format=1.1e-1] S[table-format=3.2]}
        \toprule
        {\% photon survival} & {Required depth ($t = x/X_0$)} \\ 
        \midrule
        1.0e+01 & 2.96\,$X_0$ \\
        1.0e+00 & 5.92\,$X_0$ \\
        1.0e-01 & 8.88\,$X_0$ \\
        1.0e-02 & 11.84\,$X_0$ \\
        1.0e-03 & 14.80\,$X_0$ \\
        \bottomrule
    \end{tabular}
\end{table}

\textsc{Pythia\,8} simulations of soft-QCD processes at 
$\sqrt{s} = 100\,\text{TeV}$, with the selection criteria 
$|\eta| < 4$ and $p_\text{T} > 0.5\,\text{GeV}$, yield 
approximately $10^4$ photons per bunch crossing at a pile-up 
of 1000. A suppression factor of 
$10^{-4}$ is therefore required to reduce this to 
$\mathcal{O}(1)$ photon per bunch crossing, which from 
Table~\ref{tab:photon_shielding_attenuation} corresponds to 
approximately $12\,X_0$ of shielding material.

This can be achieved with the composite shield described in 
Section\,\ref{ssec:material}. The radiation lengths of 
TZM and WCu80 in physical units are approximately $0.96\,\text{cm}$ 
and $0.43\,\text{cm}$, respectively~\cite{pdg2024}. A shield 
comprising $4\,\text{cm}$ of TZM followed by $4\,\text{cm}$ of 
WCu80 therefore accumulates a combined depth of,
\begin{equation}
    t = \frac{4\,\text{cm}}{0.96\,\text{cm}/X_0} + 
        \frac{4\,\text{cm}}{0.43\,\text{cm}/X_0} 
    \approx 4.2\,X_0 + 9.3\,X_0 \approx 13.5\,X_0,
\end{equation}
with a final $2\,\text{cm}$ of boron-loaded polymer appended 
to absorb thermal neutrons, bringing the total shield thickness 
to $10\,\text{cm}$. This configuration achieves the required 
${\sim}12\,X_0$ of electromagnetic absorption while leaving the 
hadronic flux largely unaffected, making it well-suited to 
photon-specific background rejection.

While this analytical estimate assumes primary photon incidence, hadronic interactions within the shield can repopulate the photon flux through secondary production. Consequently, we utilize a full \textsc{Geant4}-based simulation in the following section to quantify these secondaries and verify the residual background levels.

\vspace*{-0.4cm}

\subsubsection*{\textbf{Thin shield -- {\rm \textsc{Geant4}} response}}

\vspace*{-0.2cm}

In this section we present the \textsc{Geant4} simulation 
response of the thin composite shield described in the 
preceding section. The same \texttt{FTFP\_BERT} physics list is employed here.
Thin shielding response is calculated in the same manner as discussed in \ref{ssec:thickness}. The resulting spectrum from the thin shielding response is presented in Table~\ref{tab:particle_yield_summary_100TeV}.

The simulation results confirm a highly effective suppression of the electromagnetic background. The combined photon and electron flux 
from soft-QCD amounts to approximately 12000 particles per 
bunch crossing prior to the shield.
After traversing the $10\,\text{cm}$ composite shield, this is reduced to fewer than 120 photons and electrons per bunch crossing, corresponding to a 
suppression factor of $\mathcal{O}(10^{-2})$. A comparable suppression factor 
is observed in each individual energy bin. It should be noted that this residual electromagnetic flux is not composed solely of primary soft-QCD photons and electrons that have survived the shield. Rather, it receives a significant contribution from secondary photons and electrons produced via hadronic interactions of the surviving pion, kaon, and nucleon flux within the shield material itself. 
As a result, the residual electromagnetic background downstream of the shield is also coupled to the hadronic flux, and hence, cannot be eliminated entirely by a thin shield of 10\,cm.

For the hadronic component, primarily for charged pions 
($\pi^{\pm}$), the thin shield reduces the flux by approximately 
$50\%$. This is consistent with the expected attenuation for a shield of $10\,\text{cm}$ thickness, given the 
nuclear interaction lengths of TZM and WCu80 discussed in 
Section~\ref{ssec:material}. In summary, the thin 
composite shield suppresses the electromagnetic flux by a 
factor of $\mathcal{O}(100)$ while attenuating the hadronic 
flux by a factor of $\mathcal{O}(2)$, providing substantial cleaning of the ECAL whilst partially reducing the HCAL background as well.

It should be emphasized that this electromagnetic suppression 
is most effective when a tracking layer is deployed 
immediately downstream of the shield. In this configuration, 
the residual charged pion flux passing through the shield 
can be identified via track association and vetoed from the 
ECAL analysis, as discussed in next
Section.
The combination of a thin 
shielding and downstream tracking thus provides a powerful 
and complementary strategy for improving both ECAL-based LLP 
searches and broader calorimetric analyses at the FCC-hh.

Moreover, at the extreme pile-up environment of FCC-hh 
($\langle\mu\rangle \sim 1000$), ECAL performance variables such 
as timing resolution, cluster shape discrimination, and isolation 
criteria are significantly degraded by the high flux of soft 
particles. By suppressing the electromagnetic flux by O(100) and 
the hadronic flux by approximately 50\%, the thin shield substantially 
reduces this pile-up-induced contamination, enhancing both the timing 
precision and signal-to-background separation crucial for LLP 
identification.

\section{Testing the shielding strategy at the HL-LHC}
\label{sec:HL_LHC}

The discussion in the preceding sections has focused on the FCC-hh, 
which, if approved, is expected to begin operations around 2070s\,\cite{CERN_FCC_Overview}. In the nearer term, the High-Luminosity LHC (HL-LHC) is scheduled to begin 
its first physics run (Run~4) in June 2030~\cite{CERN_HLLHC_Schedule}, 
with operations planned to continue through to 2041. 
This timeline provides a unique opportunity for preparatory benchmarking and operational validation of the shielding strategies developed in this work.

We propose that a dedicated period during the later stages of HL-LHC 
operation could serve as a valuable testing ground for the shielding 
strategy described in this work, providing both physics results and 
crucial operational experience relevant to FCC-hh planning.

\vspace*{-0.4cm}

\subsection{Thick shielding at HL-LHC}
\label{ssec:thick_shield_HLLHC}

\vspace*{-0.3cm}

During the later phases of the HL-LHC run, portions of the inner tracker 
could be removed and replaced by a cylindrical shielding insert of 
20--30\,cm thickness. According to the attenuation estimates presented 
in Table~\ref{tab:shielding_attenuation}, such a configuration would 
suppress approximately 80--90\% of the incident hadron flux, 
substantially reducing the SM hadronic background while 
leaving LLPs, which interact only weakly with the shield, largely unaffected. This would 
directly enhance the LLP sensitivity of the main detector systems.

Beyond the improvement in LLP sensitivity, this configuration would 
enable a number of complementary studies --

\paragraph{Validation of anomalous muon spectrometer hits.}
As discussed in Section~\ref{ssec:improved_MS_analysis_FCC}, a 
comparison of the rate of anomalous MS hits before and after insertion of the shielding would provide a direct 
experimental handle on the nature of such events. A significant 
reduction in the anomalous hit rate upon insertion of the shield would 
identify the excess as hadronic punch-through background, whereas an 
approximately unchanged rate would constitute evidence for a genuine 
LLP contribution. Any such signals identified at the HL-LHC could 
then be studied in detail at the FCC-hh, providing direct experimental 
guidance for FCC detector design.

\paragraph{LLP models with photons in the final states.}
The shielded configuration would also be particularly sensitive to 
exotic LLP signatures with strong dependence on the presence of 
shielding. A prominent example is the class of models in 
which an LLP decays to photons. Since the shield strongly attenuates 
the prompt photon background, an observation of delayed photon 
signatures in the ECAL -- delayed 
relative to prompt production due to the finite LLP mass and flight 
time -- would constitute a robust signal for such models that would 
be very difficult to mimic with SM processes in the 
shielded environment. However, such searches may face substantial background from charged 
pions misidentified as photons in the ECAL. This necessitates the 
development of dedicated reconstruction and timing-based algorithms 
to distinguish genuine photons from hadronic deposits, as discussed 
in the next section.

\paragraph{In-situ radiation hardness testing.}
Finally, the intense particle flux of the HL-LHC provides an 
exceptional environment for material irradiation studies. The late 
stages of HL-LHC operation could serve as a testbed for assessing the 
radiation hardness and long-term mechanical performance of candidate 
shielding materials, including those proposed for the FCC-hh shield 
described in this work. Such measurements would directly inform 
material selection for the FCC-hh shielding program and may also 
find applications in broader radiation-hardness studies relevant to 
future space instrumentation and nuclear engineering.

In summary, a shielding based detector concept at the HL-LHC would serve a 
dual purpose -- delivering physics results on LLP searches in the 
near term, while providing the experimental proof-of-concept required to realize the full-scale \textsf{DELIGHT-SHIELD} program at 100\,TeV.

\vspace*{-0.4cm}

\subsection{Thin shielding at the HL-LHC}
\label{ssec:thin_shield_HLLHC}

\vspace*{-0.3cm}

Even in 
scenarios where a thick shielding insert is not operationally 
feasible at the HL-LHC, a thin $10\,\text{cm}$ composite 
shield inserted in place of the inner tracker can still 
provide meaningful background suppression. Analogous to the FCC-hh configuration, this scenario does not necessitate removing the entire CMS tracking system. Only the pixel detector modules of the inner tracker~\cite{Rossi:2816248} would be replaced, either 
completely or partially, by the thin shield assembly, leaving the outer tracker subsystems unaffected. As established in 
Section~\ref{ssec:thin_shielding_FCC}, such a shield efficiently 
attenuates the electromagnetic component of the soft-QCD 
flux while leaving the hadronic component, primarily 
charged pions ($\pi^{\pm}$), largely unaffected.

To quantify this for HL-LHC conditions, the incident particle 
spectra are generated using \textsc{Pythia\,8}~\cite{Bierlich:2022pfr} 
for soft-QCD processes at $\sqrt{s} = 14\,\text{TeV}$, with 
selection criteria $|\eta| < 1.44$, $p_\text{T} > 0.5\ \text{GeV}$, 
and a mean pile-up of $\langle\mu\rangle = 200$. The energy spectrum of each particle species is binned in $1~ \text{GeV}$ intervals, with bin centers ranging from $1$ to $15\,\text{GeV}$ and the dominant contributions are tabulated in Table~\ref{tab:softqcd_results_14TeV} in the Appendix. The 
\textsc{Geant4} shield response is then evaluated following 
the procedure described in 
Section~\ref{ssec:thin_shielding_FCC}, with secondary 
particle yields above $0.5\,\text{GeV}$ recorded and 
normalised to the primary soft-QCD multiplicities. The 
results are presented in Table~\ref{tab:particle_yield_summary_14TeV} in the Appendix. 
Consistent with the FCC-hh case, the total photon flux is 
suppressed by a factor of $\mathcal{O}(100)$, while the 
charged pion flux is reduced by approximately $50\%$, 
confirming that thin composite shielding is equally effective 
at the HL-LHC center-of-mass energy.


The insertion of the thin shield in place of the inner tracker 
introduces a distinct analysis challenge. In the absence of 
inner tracker hits, charged pion deposits in the ECAL cannot 
be straightforwardly distinguished from photon solely based on outer tracker, 
potentially degrading the purity of ECAL-based LLP analyses. 
This problem arises in both the thin-shield configuration at 
the HL-LHC and the analogous configuration at the FCC-hh, as discussed in Section\,\ref{ssec:thin_shielding_FCC}.

This challenge can be addressed by exploiting the information 
available from 6 layers of the CMS barrel outer tracker\,\cite{Rossi:2816248} and the MIP Timing 
Detector (MTD)\,\cite{CMS_MTD_TDR}. The MTD is designed to 
measure the arrival time of charged particles with a 
resolution of $30$--$60\,\text{ps}$ throughout the HL-LHC 
run~\cite{CMS_MTD_status_2026}. Since charged pions leave 
hits in the outer tracker and a timing signature in the MTD, 
whereas photons do not, a combined tracking and timing 
algorithm can be developed to tag $\pi^{\pm}$ candidates 
and veto their associated ECAL deposits prior to the LLP 
analysis. The feasibility of such algorithms has been 
demonstrated in the context of charged hadron identification 
via time-of-flight measurements at the 
HL-LHC~\cite{Cerri:2018rkm}. An algorithm developed and 
validated using HL-LHC data in the shielded configuration 
could subsequently be adapted and deployed at the FCC-hh, 
where analogous timing detector systems are anticipated, 
providing a direct and well-tested technology transfer 
pathway between the two programs.

\vspace*{-0.4cm}

\subsection{Physics analyses with a shielded interaction point}
\label{ssec:shielded_IP_physics}

\vspace*{-0.3cm}

The replacement of the inner tracker with shielding 
necessarily reduces the tracking capabilities of the 
detector. In this section we explore which physics analyses 
remain viable, and indeed are enhanced, in this configuration 
at the HL-LHC.
Three complementary analysis strategies are investigated. 
First, we study the impact of the shielding on jet-based analyses. 
Second,  we study the spectrum of particles leaking into MS as a function of the shielding thickness, which would contribute to the punch-through activity. 
Third, we briefly discuss the possibility of exploiting timing information from the MTD and energy deposits in the ECAL on LLP searches in the presence of the shielding.

\subsubsection*{\textbf{Jet analysis with shielding at the HL-LHC}}
\label{sssec:jet_analysis_HLLHC}
\vspace*{-0.2cm}

In this section we study the effect of the shielding insert on jet reconstruction in the combined ECAL and HCAL system. Soft-QCD events at $\sqrt{s} = 14\,\text{TeV}$ are first generated using \textsc{Pythia\,8} and stored in HepMC format~\cite{hepmc}. A cylindrical WCu80 shielding layer of thickness 10\,cm or $20\,\text{cm}$ is then placed around the IP in \textsc{Geant4}, followed by a segmented ECAL and HCAL geometry. The \textsc{Pythia\,8} output is passed through the \textsc{Geant4} simulation and the resulting energy deposits are recorded as hits in the calorimeter cells. The \texttt{FTFP\_BERT} physics list is also employed here. To simulate a pile-up of $\langle\mu\rangle = 200$, the \textsc{Geant4} output of 200 randomly selected soft-QCD events is overlaid, with the total energy deposited per calorimeter cell summed across 
all events. 
Jets are reconstructed from the calorimeter cell energies using 
the anti-$k_t$ algorithm~\cite{Cacciari:2008gp} with a distance 
parameter of $R = 0.4$ and a minimum jet $p_\text{T}$ of 
$p_\text{T}^\text{jet} > 20\,\text{GeV}$, as implemented via 
the \texttt{pyjet} Python interface~\cite{pyjet_zenodo} to the 
\textsc{FastJet} package~\cite{Cacciari:2011ma}. 
The full procedure is repeated independently for the three shielding configurations -- no shield, and $10\,\text{cm}$ and $20\,\text{cm}$ of WCu80.

\begin{figure}[h]
\centering
\includegraphics[width=0.7\linewidth]{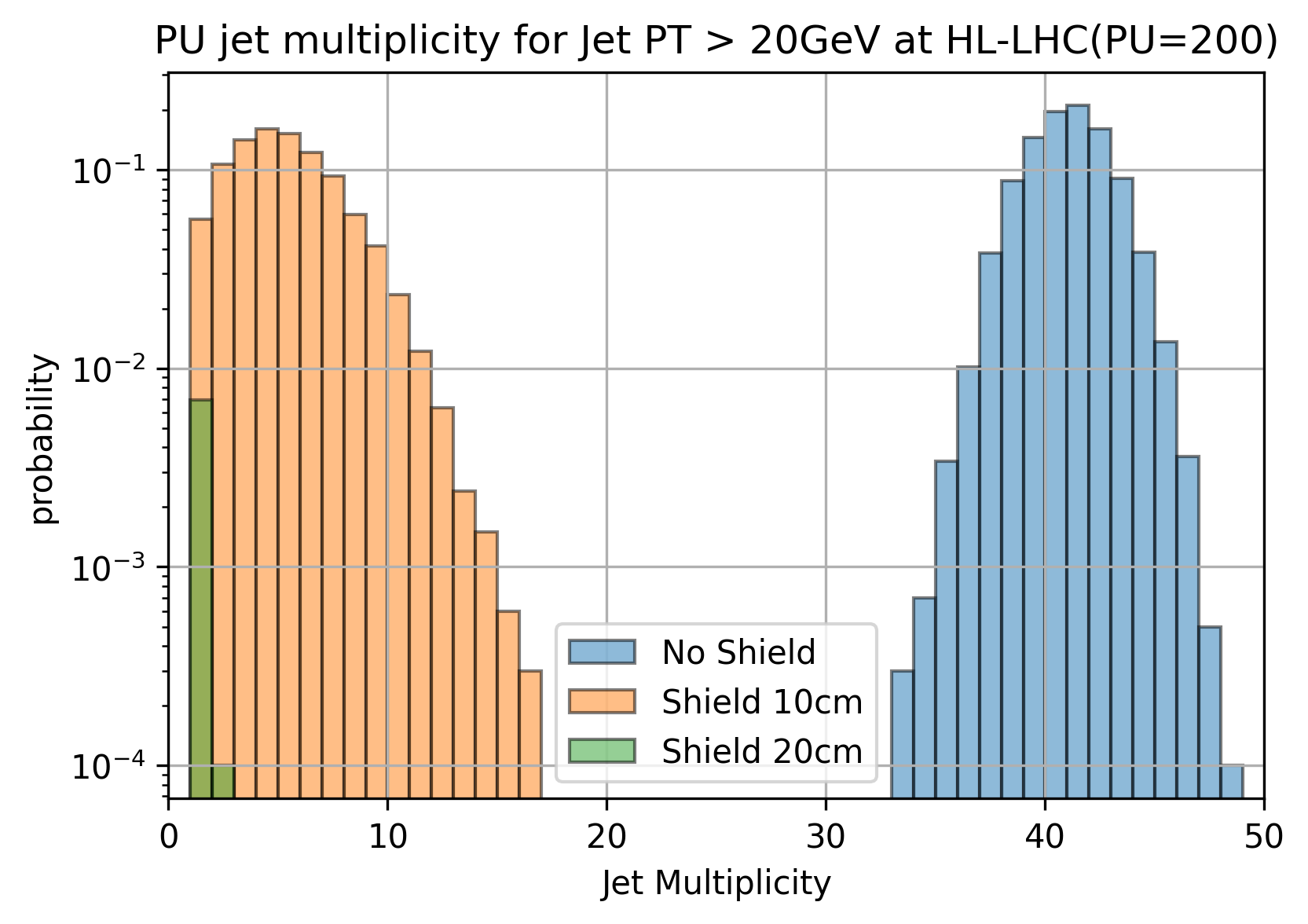}
\caption{Jet multiplicity distribution per bunch crossing at 
$\sqrt{s} = 14\,\text{TeV}$ with $\langle\mu\rangle = 200$ 
pile-up, for three shielding configurations -- no shielding, $10\,\text{cm}$ WCu80, and $20\,\text{cm}$ WCu80. 
Jets are clustered using the anti-$k_t$ algorithm with 
$R = 0.4$ and $p_\text{T}^\text{jet} > 20\,\text{GeV}$.}
\label{jet_mult}
\end{figure}

The resulting jet multiplicity distributions per bunch crossing 
are shown in Fig.\,\ref{jet_mult}. Without shielding, the 
200 PU scenario produces approximately 40--45 reconstructed 
jets per bunch crossing , with every 
bunch crossing containing at least one reconstructed 
jet with $p_{\rm T}>20$GeV. Insertion of $10\,\text{cm}$ of WCu80 
shielding reduces this to 4--5 jets per bunch crossing with $98\%$ of bunch crossings still 
containing at least one jet with $p_{\rm T}>20$GeV. Further insertion of  
$20\,\text{cm}$ of WCu80 reduces the multiplicity to 
1--2 jets per bunch crossing, with only $0.7\%$ of 
bunch crossings containing any reconstructed jet with $p_{\rm T}>20$GeV at all. This demonstrates a strong and monotonic dependence of the reconstructed pile-up jet multiplicity and the probability of a collision event to have at least one reconstructed jet, on the shield thickness.
The jet $p_\text{T}$ distributions for the three 
configurations are shown in Fig.~\ref{jet_PT_hist} in the Appendix. 
In addition to the reduction in jet multiplicity, the shielding 
significantly reduces the maximum reconstructed jet $p_\text{T}$ -- 
without shielding, jets are reconstructed up to 
$p_\text{T} \approx 150\,\text{GeV}$ from soft-QCD, whereas insertion of 
$10\,\text{cm}$ and 20\,cm of WCu80 reduces this to below 
$50\,\text{GeV}$ and 30\,GeV, respectively.

Taken together, these results demonstrate that the shielding component produces a clear and quantifiable reduction in both the multiplicity and the $p_T$ spectrum of PU jets in the calorimeter. It efficiently suppresses low-energy jets from soft-QCD processes in high PU interactions, a dominant background for hadronic LLP decays. The remaining jet activity can be further reduced by raising the minimum jet $p_T$ threshold, exploiting the softer spectrum emanating after the shielded configuration. Overall, the shielding cleans the calorimetric environment, improves the signal-to-background ratio, and enhances a broad class of LLP searches at the HL-LHC.

\vspace*{-0.4cm}
\subsubsection*{\textbf{Punch-through jets in the MS analysis}}

\vspace*{-0.2cm}
Punch-through jets refer to hadronic jets whose particles are not fully absorbed in the calorimeter system and instead penetrate into the muon detectors. Punch-through jets produce additional hits that complicate muon reconstruction and contribute to background in searches involving displaced muons or non-prompt hadronic activities.
In this section, we investigate the effect of the extra shielding component on the punch-through activity by using \textsc{Geant4} simulations.
To assess the punch-through suppression provided by the extra shielding around the IP, we first generate hard-QCD events using \textsc{Pythia8} with a minimum transverse momentum threshold of \qty{1}{\TeV}. The generated particles are then propagated through a \textsc{Geant4} simulation of the HL-LHC ECAL and HCAL, where the distributions of particles exiting the HCAL are recorded. 
This process is repeated with the addition of two different shielding configurations -- a \qty{60}{\cm} shield (\qty{25}{\cm} TZM + \qty{25}{\cm} WCu80 + \qty{10}{\cm} borated layer) and a \qty{110}{\cm} shield (\qty{50}{\cm} TZM + \qty{50}{\cm} WCu80 + \qty{10}{\cm} borated layer). A \qty{1}{\GeV} kinetic energy threshold is applied to the exiting particles. The flux suppression factor is then calculated as the ratio of the particle flux with the \qty{60}{\cm} or \qty{110}{\cm} shielding to the baseline flux without the extra shielding. The results are summarized in Table~\ref{tab:MS_flux_suppression}.
\begin{table}[htbp]
\centering
\caption{HardQCD at $\sqrt{s}=$14\,TeV (with a $p_T$ threshold of 1\,TeV): flux suppression factors for \qty{60}{\cm} and \qty{110}{\cm} thick shields relative to the no-shielding baseline.}
\begin{tabular}{l S[table-format=1.3] S[table-format=1.3]}
\hline
\textbf{Particle} & {\textbf{60 cm}} & {\textbf{110 cm}} \\
\hline
$e^{\pm}$      & 2.427 & 3.636 \\
$\gamma$       & 3.311 & 3.861 \\
p              & 3.831 & 4.695 \\
n              & 4.132 & 6.024 \\
$\pi^{\pm}$    & 4.132 & 6.579 \\
$K^{\pm}$      & 4.630 & 7.692 \\
$K^0$          & 4.000 & 8.621 \\
$\mu^{\pm}$    & 2.294 & 2.525 \\
\hline
\end{tabular}
\label{tab:MS_flux_suppression}
\end{table}

These results clearly demonstrate that a \qty{60}{\cm} shield provides a suppression factor of approximately 4 for the hadronic flux, while the \qty{110}{\cm} configuration yields a suppression factor between 6 and 8. Similarly, the electromagnetic flux is suppressed by a factor of roughly 2.5 to 3.3 with the \qty{60}{\cm} shield, increasing to nearly a factor of 4 with the \qty{110}{\cm} shielding.



\vspace*{-0.4cm}

\subsubsection*{\textbf{Timing and energy cut analysis}}

\vspace*{-0.2cm}

The shielding could also potentially increase the prospect of the timing-based LLP searches.
If the timing layer is placed downstream of the shielding, a major part of the PU activity is expected to get suppressed. In Refs.\,\cite{Bhattacherjee:2020nno,Bhattacherjee:2021qaa}, it was shown how PU adversely affects jet timing observables.
Moreover, a non-negligible contribution from SM LLPs, such as $K_S$, was also observed.
If the shielding is able to suppress these backgrounds, we expect an improvement in jet timing observables as well.
A quantitative assessment of this effect, however, requires a detailed treatment of the $p_T$- and $\eta$-dependent timing resolution, its degradation over HL-LHC running, and the spatial and temporal spread of PU vertices. Such a study is beyond the scope of the present paper and is left for future investigation.


\vspace*{-0.4cm}

\section{Summary and Outlook}
\label{sec:summary}

\vspace*{-0.3cm}

In this work, we have proposed and investigated the concept of a 
dedicated IP for LLP searches at 
the FCC-hh, equipped with a cylindrical composite shielding 
structure designed to suppress SM backgrounds arising 
primarily from soft-QCD processes. We have further proposed a 
novel detector concept, \textsf{DELIGHT-SHIELD}, to be deployed at this 
IP, and demonstrated its potential sensitivity 
across a broad range of LLP models and parameter space.

The shielding design is proposed to comprise of three material layers -- TZM, 
WCu80, and a boron-loaded polymer composite -- where TZM and WCu80 provide complementary contributions to hadronic attenuation and electromagnetic suppression, while the boron-loaded polymer composite is optimized for thermal neutron capture. Analytical estimates and full 
\textsc{Geant4} simulations confirm that a total shield thickness of 
approximately $1.5$--$2\,\text{m}$ suppresses the soft-QCD 
hadronic flux by a factor of $10^{-5}$--$10^{-7}$, sufficient 
to reduce the incident flux of $\mathcal{O}(10^4)$ hadrons per 
bunch crossing to well below one penetrating particle on average. 
A preliminary thermal analysis demonstrates that the heat load deposited by the soft-QCD particle flux would not hamper the structural integrity of the shielding.

Sensitivity projections for the benchmark processes 
$h \rightarrow \phi\phi$, $B \rightarrow \phi K$, and 
$D/D_s \rightarrow N\mu$ demonstrate that \textsf{DELIGHT-SHIELD} 
achieves substantially improved sensitivity relative to the 
FCC-hh barrel MS across all LLP model 
categories studied. For the $h \rightarrow \phi\phi$ process, 
certain geometric configurations achieve sensitivity to 
branching ratios as low as $\mathcal{O}(10^{-9})$ at $c\tau \approx 1\,\text{m}$, providing a high gain in sensitivity as compared to the MS based LLP searches.
Conversely, even if the dedicated IP is operated for only a fraction of the full FCC-hh integrated luminosity, \textsf{DELIGHT-SHIELD} can match or surpass 
the full-luminosity MS sensitivity. A shield-in/shield-out mechanism can further help in identifying the nature of anomalous MS activity, in case an excess is observed. Furthermore, we demonstrate that a relatively thin shielding can reduce the EM background for ECAL based LLP searches. Beyond the FCC-hh, we have proposed several complementary 
applications and testing of the shielding concept at the HL-LHC.

In conclusion, the dedicated shielded IP represents a highly efficient expansion of the FCC-hh physics program. 
By integrating near-term material qualification and physics benchmarking at the HL-LHC, the \textsf{DELIGHT-SHIELD} program offers a clear and robust path toward exploring the lifetime frontier at the next generation of hadron colliders.

\vspace{0.5cm}


\textbf{\textit{Acknowledgments}} -- 
 The work of BB is  supported by the Core Research Grant CRG/2022/001922 of the Science and Engineering Research Board (SERB), Government of India. BB and AS are grateful to the Center for High Energy Physics at the Indian Institute of Science for the cluster facility. 
The work of SM is supported by an initiation grant \texttt{(IITK/PHY/2023282)} received from IIT Kanpur. AC acknowledges the financial support provided by the same initiation grant \texttt{(IITK/PHY/2023282)} 
during his three-month tenure as a project student at IIT Kanpur from January to March 2026.
\bibliographystyle{utphys}
\bibliography{reference}

\newpage
\appendix
\section{}

\begin{table}[htbp]
\centering
\caption{Ratio of particle yields  with $E > 1\,\text{GeV}$, for C26000 relative to WCu80 under pion bombardment at different energies and shield thicknesses, for simulated 1000 events. A value greater than 1 indicates higher secondary particle production in C26000 compared to WCu80. Entries marked with ``--'' correspond to undefined ratios due to zero yield in WCu80.}
\label{tab:brass_W_yield_ratio}
\normalsize
\setlength{\tabcolsep}{4pt}
\begin{tabular}{cccccccccc}
\toprule
{$E$ [GeV]} & {Depth [cm]} & $N_\text{tot}$ & $e/\gamma$ & $\pi^{\pm}$ & $p$ & $n$ & $K^{\pm}$ & $K^0$ & $\mu$ \\
\midrule
\multirow{3}{*}{50} 
 & 50  & 1.74 & 4.16 & 1.68 & 1.60 & 1.38 & 1.72 & 1.47 & 4.00 \\
 & 100 & 3.75 & --   & 3.60 & 3.58 & 2.26 & 4.00 & 12.00 & 4.00 \\
 & 200 & 3.00 & --   & --   & --   & --   & --   & --   & --   \\
\multirow{3}{*}{100} 
 & 50  & 1.63 & 3.10 & 1.56 & 1.52 & 1.35 & 1.65 & 1.46 & 4.00 \\
 & 100 & 2.81 & 5.36 & 2.93 & 2.33 & 2.20 & 3.64 & 2.64 & 2.67 \\
 & 200 & 1.67 & 0.33 & 1.00 & --   & --   & --   & --   & --   \\
\multirow{3}{*}{500} 
 & 50  & 1.47 & 3.04 & 1.29 & 1.31 & 1.22 & 1.54 & 1.44 & 1.47 \\
 & 100 & 2.41 & 6.38 & 2.22 & 2.21 & 2.38 & 2.59 & 2.19 & 2.25 \\
 & 200 & 4.89 & 9.00 & 6.36 & 22.00& 15.00& --   & --   & 1.00 \\
\bottomrule
\end{tabular}
\end{table}

\begin{table}[htbp]
    \centering
    \caption{Secondary particle yields per $10^5$ primary $\gamma$ particles 
             with $E > 0.5\,\text{GeV}$, for total shield thicknesses of 60\,cm, 110\,cm, 
             and 210\,cm.}
    \label{tab:thick_geant4_gamma}
    \normalsize
    \setlength{\tabcolsep}{4pt}
    \begin{tabular}{cc rrrrrrrr}
        \toprule
        {$E$ [GeV]} & {Depth [cm]} & $N_\text{tot}$ & $e/\gamma$ & $\pi^{\pm}$ & $p$ & $n$ & $K^{\pm}$ & $K^0$ & $\mu$ \\
        \midrule
        \multirow{3}{*}{10} & 50 & 15 & 0 & 6 & 4 & 4 & 0 & 0 & 1 \\
         & 100 & 1 & 0 & 1 & 0 & 0 & 0 & 0 & 0 \\
         & 200 & 0 & 0 & 0 & 0 & 0 & 0 & 0 & 0 \\
        \addlinespace
        \multirow{3}{*}{20} & 50 & 39 & 0 & 13 & 7 & 18 & 0 & 0 & 0 \\
         & 100 & 4 & 0 & 2 & 0 & 2 & 0 & 0 & 0 \\
         & 200 & 0 & 0 & 0 & 0 & 0 & 0 & 0 & 0 \\
        \addlinespace
        \multirow{3}{*}{50} & 50 & 75 & 0 & 27 & 16 & 30 & 1 & 1 & 0 \\
         & 100 & 6 & 0 & 3 & 0 & 2 & 0 & 0 & 0 \\
         & 200 & 0 & 0 & 0 & 0 & 0 & 0 & 0 & 0 \\
        \addlinespace
        \multirow{3}{*}{100} & 50 & 248 & 18 & 94 & 38 & 82 & 9 & 5 & 2 \\
         & 100 & 11 & 0 & 7 & 2 & 2 & 0 & 0 & 0 \\
         & 200 & 0 & 0 & 0 & 0 & 0 & 0 & 0 & 0 \\
        \addlinespace
        \multirow{3}{*}{200} & 50 & 618 & 19 & 319 & 70 & 172 & 14 & 15 & 4 \\
         & 100 & 55 & 7 & 20 & 4 & 16 & 4 & 2 & 1 \\
         & 200 & 2 & 0 & 0 & 0 & 0 & 0 & 0 & 0 \\
        \addlinespace
        \bottomrule
    \end{tabular}
\end{table}

\begin{table}[htbp]
    \centering
    \caption{Secondary particle yields per $10^5$ primary $e^-$ particles 
             with $E > 0.5\,\text{GeV}$, for total shield thicknesses of 60\,cm, 110\,cm, 
             and 210\,cm.}
    \label{tab:thick_geant4_em}
    \normalsize
    \setlength{\tabcolsep}{4pt}
    \begin{tabular}{cc rrrrrrrr}
        \toprule
        {$E$ [GeV]} & {Depth [cm]} & $N_\text{tot}$ & $e/\gamma$ & $\pi^{\pm}$ & $p$ & $n$ & $K^{\pm}$ & $K^0$ & $\mu$ \\
        \midrule
        \multirow{3}{*}{10} & 50 & 6 & 0 & 1 & 2 & 3 & 0 & 0 & 0 \\
         & 100 & 0 & 0 & 0 & 0 & 0 & 0 & 0 & 0 \\
         & 200 & 0 & 0 & 0 & 0 & 0 & 0 & 0 & 0 \\
        \addlinespace
        \multirow{3}{*}{20} & 50 & 15 & 0 & 4 & 3 & 7 & 1 & 0 & 0 \\
         & 100 & 0 & 0 & 0 & 0 & 0 & 0 & 0 & 0 \\
         & 200 & 0 & 0 & 0 & 0 & 0 & 0 & 0 & 0 \\
        \addlinespace
        \multirow{3}{*}{50} & 50 & 52 & 0 & 15 & 9 & 23 & 2 & 1 & 1 \\
         & 100 & 1 & 0 & 0 & 0 & 1 & 0 & 0 & 0 \\
         & 200 & 0 & 0 & 0 & 0 & 0 & 0 & 0 & 0 \\
        \addlinespace
        \multirow{3}{*}{100} & 50 & 172 & 6 & 71 & 25 & 61 & 5 & 0 & 2 \\
         & 100 & 5 & 0 & 4 & 1 & 0 & 0 & 0 & 0 \\
         & 200 & 0 & 0 & 0 & 0 & 0 & 0 & 0 & 0 \\
        \addlinespace
        \multirow{3}{*}{200} & 50 & 462 & 15 & 220 & 60 & 145 & 7 & 7 & 2 \\
         & 100 & 23 & 0 & 12 & 1 & 7 & 0 & 0 & 2 \\
         & 200 & 4 & 0 & 0 & 0 & 0 & 0 & 0 & 1 \\
        \addlinespace
        \bottomrule
    \end{tabular}
\end{table}

\begin{table}[htbp]
    \centering
    \caption{Secondary particle yields per $10^5$ primary $\pi^{\pm}$ particles 
             with $E > 0.5\,\text{GeV}$, for total shield thicknesses of 60\,cm, 110\,cm, 
             and 210\,cm.}
    \label{tab:thick_geant4_pip}
    \normalsize
    \setlength{\tabcolsep}{4pt}
    \begin{tabular}{cc rrrrrrrr}
        \toprule
        {$E$ [GeV]} & {Depth [cm]} & $N_\text{tot}$ & $e/\gamma$ & $\pi^{\pm}$ & $p$ & $n$ & $K^{\pm}$ & $K^0$ & $\mu$ \\
        \midrule
        \multirow{3}{*}{10} & 50 & 31,814 & 1,535 & 19,536 & 3,110 & 6,050 & 668 & 458 & 225 \\
         & 100 & 1,966 & 76 & 897 & 149 & 419 & 28 & 24 & 165 \\
         & 200 & 290 & 0 & 0 & 0 & 1 & 0 & 0 & 137 \\
        \addlinespace
        \multirow{3}{*}{20} & 50 & 78,904 & 4,641 & 46,186 & 8,749 & 14,879 & 2,138 & 1,553 & 314 \\
         & 100 & 5,865 & 257 & 2,939 & 660 & 1,357 & 150 & 77 & 170 \\
         & 200 & 238 & 0 & 4 & 2 & 3 & 0 & 0 & 80 \\
        \addlinespace
        \multirow{3}{*}{50} & 50 & 222,575 & 16,972 & 126,485 & 25,023 & 39,207 & 7,492 & 5,159 & 756 \\
         & 100 & 21,787 & 1,157 & 11,222 & 2,467 & 4,926 & 655 & 457 & 326 \\
         & 200 & 519 & 1 & 45 & 6 & 25 & 1 & 2 & 101 \\
        \addlinespace
        \multirow{3}{*}{100} & 50 & 462,848 & 41,852 & 259,801 & 50,631 & 77,944 & 16,484 & 11,602 & 1,249 \\
         & 100 & 55,276 & 3,480 & 28,735 & 6,375 & 11,892 & 1,881 & 1,209 & 565 \\
         & 200 & 974 & 14 & 131 & 32 & 58 & 11 & 6 & 173 \\
        \addlinespace
        \multirow{3}{*}{200} & 50 & 934,670 & 98,163 & 518,142 & 100,217 & 150,283 & 34,246 & 24,369 & 2,189 \\
         & 100 & 132,445 & 8,758 & 69,454 & 15,532 & 27,679 & 4,488 & 3,077 & 1,155 \\
         & 200 & 2,127 & 49 & 428 & 105 & 215 & 18 & 14 & 303 \\
        \addlinespace
        \bottomrule
    \end{tabular}
\end{table}

\begin{table}[htbp]
    \centering
    \caption{Secondary particle yields per $10^5$ primary $p$ particles 
             with $E > 0.5\,\text{GeV}$, for total shield thicknesses of 60\,cm, 110\,cm, 
             and 210\,cm.}
    \label{tab:thick_geant4_proton}
    \normalsize
    \setlength{\tabcolsep}{4pt}
    \begin{tabular}{cc rrrrrrrr}
        \toprule
        {$E$ [GeV]} & {Depth [cm]} & $N_\text{tot}$ & $e/\gamma$ & $\pi^{\pm}$ & $p$ & $n$ & $K^{\pm}$ & $K^0$ & $\mu$ \\
        \midrule
        \multirow{3}{*}{10} & 50 & 34,428 & 865 & 7,905 & 12,265 & 13,028 & 133 & 87 & 46 \\
         & 100 & 1,823 & 35 & 337 & 572 & 837 & 3 & 1 & 8 \\
         & 200 & 17 & 0 & 0 & 1 & 1 & 0 & 0 & 0 \\
        \addlinespace
        \multirow{3}{*}{20} & 50 & 80,878 & 2,942 & 29,439 & 21,016 & 25,255 & 964 & 636 & 201 \\
         & 100 & 5,118 & 124 & 1,438 & 1,281 & 2,007 & 32 & 41 & 51 \\
         & 200 & 90 & 0 & 7 & 5 & 9 & 0 & 0 & 6 \\
        \addlinespace
        \multirow{3}{*}{50} & 50 & 227,566 & 12,899 & 105,831 & 42,054 & 55,402 & 5,306 & 3,678 & 646 \\
         & 100 & 19,223 & 687 & 7,711 & 3,582 & 5,925 & 364 & 224 & 241 \\
         & 200 & 339 & 1 & 7 & 9 & 21 & 1 & 0 & 55 \\
        \addlinespace
        \multirow{3}{*}{100} & 50 & 481,537 & 35,291 & 243,149 & 72,859 & 101,133 & 14,196 & 9,592 & 1,252 \\
         & 100 & 48,592 & 2,230 & 22,053 & 7,650 & 13,008 & 1,223 & 760 & 550 \\
         & 200 & 891 & 9 & 73 & 36 & 59 & 6 & 4 & 138 \\
        \addlinespace
        \multirow{3}{*}{200} & 50 & 980,279 & 86,808 & 516,588 & 128,644 & 182,501 & 32,522 & 22,372 & 2,264 \\
         & 100 & 122,262 & 6,939 & 59,405 & 17,023 & 29,590 & 3,536 & 2,462 & 1,093 \\
         & 200 & 1,770 & 15 & 201 & 68 & 148 & 14 & 5 & 255 \\
        \addlinespace
        \bottomrule
    \end{tabular}
\end{table}

\begin{table}[htbp]
    \centering
    \caption{Secondary particle yields per $10^5$ primary $n$ particles 
             with $E > 0.5\,\text{GeV}$, for total shield thicknesses of 60\,cm, 110\,cm, 
             and 210\,cm.}
    \label{tab:thick_geant4_neutron}
    \normalsize
    \setlength{\tabcolsep}{4pt}
    \begin{tabular}{cc rrrrrrrr}
        \toprule
        {$E$ [GeV]} & {Depth [cm]} & $N_\text{tot}$ & $e/\gamma$ & $\pi^{\pm}$ & $p$ & $n$ & $K^{\pm}$ & $K^0$ & $\mu$ \\
        \midrule
        \multirow{3}{*}{10} & 50 & 36,203 & 886 & 8,901 & 9,409 & 16,544 & 146 & 120 & 71 \\
         & 100 & 1,916 & 25 & 322 & 491 & 1,023 & 7 & 3 & 8 \\
         & 200 & 9 & 0 & 0 & 1 & 1 & 0 & 0 & 0 \\
        \addlinespace
        \multirow{3}{*}{20} & 50 & 83,031 & 3,203 & 30,514 & 18,022 & 28,987 & 951 & 677 & 209 \\
         & 100 & 5,463 & 150 & 1,535 & 1,284 & 2,235 & 37 & 23 & 50 \\
         & 200 & 70 & 1 & 4 & 2 & 4 & 0 & 0 & 6 \\
        \addlinespace
        \multirow{3}{*}{50} & 50 & 229,815 & 13,038 & 106,740 & 38,720 & 59,896 & 5,374 & 3,648 & 628 \\
         & 100 & 19,656 & 742 & 7,781 & 3,558 & 6,241 & 381 & 234 & 238 \\
         & 200 & 328 & 0 & 11 & 6 & 13 & 2 & 0 & 43 \\
        \addlinespace
        \multirow{3}{*}{100} & 50 & 480,779 & 34,771 & 242,792 & 69,750 & 104,448 & 14,014 & 9,746 & 1,181 \\
         & 100 & 49,824 & 2,342 & 22,560 & 7,707 & 13,385 & 1,243 & 837 & 566 \\
         & 200 & 779 & 2 & 56 & 23 & 38 & 2 & 4 & 143 \\
        \addlinespace
        \multirow{3}{*}{200} & 50 & 981,689 & 86,527 & 516,692 & 126,196 & 186,362 & 32,086 & 23,058 & 2,311 \\
         & 100 & 123,531 & 6,878 & 60,335 & 17,087 & 29,600 & 3,618 & 2,434 & 1,150 \\
         & 200 & 1,885 & 18 & 230 & 70 & 145 & 12 & 13 & 299 \\
        \addlinespace
        \bottomrule
    \end{tabular}
\end{table}

\begin{table}[htbp]
    \centering
    \caption{Secondary particle yields per $10^5$ primary $K^{\pm}$ particles 
             with $E > 0.5\,\text{GeV}$, for total shield thicknesses of 60\,cm, 110\,cm, 
             and 210\,cm.}
    \label{tab:thick_geant4_kaonp}
    \normalsize
    \setlength{\tabcolsep}{4pt}
    \begin{tabular}{cc rrrrrrrr}
        \toprule
        {$E$ [GeV]} & {Depth [cm]} & $N_\text{tot}$ & $e/\gamma$ & $\pi^{\pm}$ & $p$ & $n$ & $K^{\pm}$ & $K^0$ & $\mu$ \\
        \midrule
        \multirow{3}{*}{10} & 50 & 37,287 & 1,354 & 13,161 & 3,746 & 7,396 & 6,855 & 3,113 & 709 \\
         & 100 & 3,095 & 45 & 539 & 190 & 512 & 316 & 143 & 520 \\
         & 200 & 962 & 1 & 0 & 0 & 0 & 0 & 0 & 326 \\
        \addlinespace
        \multirow{3}{*}{20} & 50 & 84,097 & 3,742 & 38,158 & 9,639 & 16,249 & 9,701 & 5,002 & 638 \\
         & 100 & 5,875 & 143 & 1,970 & 593 & 1,323 & 471 & 245 & 427 \\
         & 200 & 680 & 1 & 1 & 0 & 1 & 0 & 0 & 246 \\
        \addlinespace
        \multirow{3}{*}{50} & 50 & 224,952 & 14,042 & 115,173 & 25,831 & 41,598 & 16,041 & 9,408 & 907 \\
         & 100 & 19,014 & 807 & 8,600 & 2,220 & 4,444 & 1,057 & 574 & 454 \\
         & 200 & 768 & 1 & 21 & 6 & 12 & 3 & 1 & 201 \\
        \addlinespace
        \multirow{3}{*}{100} & 50 & 465,112 & 36,090 & 247,760 & 52,117 & 82,098 & 25,499 & 16,344 & 1,361 \\
         & 100 & 46,661 & 2,306 & 22,859 & 5,453 & 10,664 & 2,044 & 1,272 & 760 \\
         & 200 & 1,167 & 9 & 69 & 17 & 43 & 3 & 2 & 243 \\
        \addlinespace
        \multirow{3}{*}{200} & 50 & 942,297 & 85,188 & 511,472 & 103,941 & 157,855 & 43,810 & 30,035 & 2,402 \\
         & 100 & 113,623 & 6,629 & 57,799 & 13,532 & 24,940 & 4,279 & 2,889 & 1,241 \\
         & 200 & 2,027 & 24 & 233 & 54 & 122 & 14 & 7 & 374 \\
        \addlinespace
        \bottomrule
    \end{tabular}
\end{table}

\begin{table}[htbp]
    \centering
    \caption{Secondary particle yields per $10^5$ primary $\mu^-$ particles 
             with $E > 0.5\,\text{GeV}$, for total shield thicknesses of 60\,cm, 110\,cm, 
             and 210\,cm. The muon survival fraction remains consistent with 
             unity at all depths and energies, confirming that muons are 
             effectively unattenuated by the shield.}
    \label{tab:thick_geant4_mum}
    \normalsize
    \setlength{\tabcolsep}{4pt}
    \begin{tabular}{cc rrrrrrrr}
        \toprule
        {$E$ [GeV]} & {Depth [cm]} & $N_\text{tot}$ & $e/\gamma$ & $\pi^{\pm}$ & $p$ & $n$ & $K^{\pm}$ & $K^0$ & $\mu$ \\
        \midrule
        \multirow{3}{*}{10} & 50  & 100,224 & 202   & 4   & 6  & 13 & 0 & 0 & 99,999 \\
         & 100 & 100,154 & 166   & 5   & 3  & 2  & 0 & 0 & 99,974 \\
         & 200 & 100,007 & 90    & 0   & 1  & 4  & 0 & 0 & 99,897 \\
        \addlinespace
        \multirow{3}{*}{20} & 50  & 100,412 & 382   & 9   & 7  & 16 & 0 & 0 & 99,998 \\
         & 100 & 100,419 & 390   & 11  & 10 & 9  & 1 & 1 & 99,995 \\
         & 200 & 100,352 & 363   & 7   & 7  & 8  & 0 & 0 & 99,960 \\
        \addlinespace
        \multirow{3}{*}{50} & 50  & 100,889 & 804   & 40  & 21 & 20 & 0 & 2 & 100,000 \\
         & 100 & 100,835 & 736   & 44  & 23 & 35 & 1 & 0 & 99,994 \\
         & 200 & 100,773 & 713   & 34  & 12 & 20 & 0 & 2 & 99,988 \\
        \addlinespace
        \multirow{3}{*}{100} & 50  & 101,844 & 1,655 & 105 & 28 & 45 & 9 & 3 & 99,997 \\
         & 100 & 101,818 & 1,607 & 113 & 33 & 48 & 7 & 5 & 99,996 \\
         & 200 & 101,494 & 1,343 & 79  & 25 & 39 & 6 & 4 & 99,998 \\
        \addlinespace
        \multirow{3}{*}{200} & 50  & 103,303 & 2,956 & 181 & 63 & 77 & 15 & 7 & 100,000 \\
         & 100 & 103,221 & 2,869 & 199 & 53 & 83 & 2  & 9 & 100,003 \\
         & 200 & 103,318 & 2,992 & 186 & 48 & 73 & 8  & 5 & 100,000 \\
        \addlinespace
        \bottomrule
    \end{tabular}
\end{table}
\begin{table}[htbp]
    \centering
    \caption{Particles spectra per bunch crossing from soft-QCD@100TeV ($|\eta| < 4$, $P_T>0.5$ GeV and $N_{\mathrm{PU}}=1000$)}
    \label{tab:softqcd_results_100TeV}
    \normalsize
    \setlength{\tabcolsep}{4pt}
    \begin{tabular}{l l l l l l l l l l}
        \toprule
        $E_{min}$ & $E_{max}$ & $e/\gamma$ & $\mu$ & $\pi^{\pm}$ & $K^{\pm}$ & $K_{L}$ & $K_{S}$ & $n$ & $p$ \\ 
        \midrule
        0.5 & 1.5 & 10829.14 & 25.10 & 19805.96 & 3177.78 & 1566.18 & 1569.02 & 1840.08 & 1841.52 \\
        1.5 & 2.5 & 1257.76 & 7.22 & 3314.70 & 835.58 & 415.24 & 414.98 & 730.18 & 730.96 \\
        2.5 & 3.5 & 275.56 & 2.68 & 881.38 & 259.46 & 129.58 & 129.46 & 251.64 & 252.88 \\
        3.5 & 4.5 & 80.56 & 1.20 & 283.28 & 91.54 & 45.30 & 45.52 & 84.92 & 84.94 \\
        4.5 & 5.5 & 29.80 & 0.58 & 106.62 & 37.22 & 18.46 & 18.60 & 31.18 & 31.30 \\
        5.5 & 6.5 & 13.06 & 0.34 & 46.62 & 17.10 & 8.64 & 8.52 & 12.90 & 12.98 \\
        6.5 & 7.5 & 6.56 & 0.20 & 23.00 & 8.86 & 4.50 & 4.38 & 6.10 & 6.08 \\
        7.5 & 8.5 & 3.52 & 0.12 & 12.40 & 5.00 & 2.42 & 2.46 & 3.24 & 3.14 \\
        8.5 & 9.5 & 2.12 & 0.06 & 7.26 & 2.88 & 1.44 & 1.54 & 1.88 & 1.84 \\
        9.5 & 10.5 & 1.24 & 0.06 & 4.62 & 1.86 & 0.90 & 0.92 & 1.08 & 1.10 \\
        10.5 & 11.5 & 0.88 & 0.02 & 2.92 & 1.24 & 0.60 & 0.62 & 0.74 & 0.74 \\
        11.5 & 12.5 & 0.60 & 0.02 & 1.90 & 0.84 & 0.42 & 0.42 & 0.46 & 0.48 \\
        12.5 & 13.5 & 0.40 & 0.02 & 1.34 & 0.58 & 0.28 & 0.30 & 0.30 & 0.32 \\
        13.5 & 14.5 & 0.30 & 0.02 & 0.98 & 0.40 & 0.22 & 0.22 & 0.22 & 0.26 \\
        14.5 & 15.5 & 0.22 & 0.02 & 0.68 & 0.32 & 0.16 & 0.14 & 0.16 & 0.16 \\
        15.5 & 16.5 & 0.18 & 0.00 & 0.52 & 0.22 & 0.14 & 0.12 & 0.12 & 0.12 \\
        16.5 & 17.5 & 0.12 & 0.00 & 0.42 & 0.16 & 0.08 & 0.08 & 0.10 & 0.10 \\
        17.5 & 18.5 & 0.08 & 0.00 & 0.30 & 0.14 & 0.06 & 0.08 & 0.06 & 0.06 \\
        18.5 & 19.5 & 0.06 & 0.00 & 0.22 & 0.12 & 0.04 & 0.04 & 0.06 & 0.04 \\
        19.5 & 20.5 & 0.06 & 0.00 & 0.22 & 0.08 & 0.04 & 0.02 & 0.02 & 0.04 \\
        \bottomrule
    \end{tabular}
\end{table}

\begin{sidewaystable}[htbp]
    \centering
    \caption{Secondary particle yields per bunch crossing (Pileup=1000) for primary soft-QCD particles
             with $E > 0.1\,\text{GeV}$, for a thick shield (thickness of 60\,cm) at 100\,TeV.}
    \label{tab:50cm_thick_shield_reponse_100TeV}
    \normalsize
    \setlength{\tabcolsep}{6pt}
    \begin{tabular}{c l l l l l l l l l l l}
        \toprule
        Energy[GeV] & $n_{tot}$ & $e/\gamma$ & $\pi^{\pm}$ & $p$ & $n$ & $K^{\pm}$ & $K^0$ & Nucleus & $\mu$ & $\nu$ & Hyperons \\ 
        \midrule
        1 & 2120.74 & 20.24 & 148.68 & 48.48 & 1261.90 & 3.46 & 32.30 & 0.00 & 84.68 & 520.08 & 0.94 \\
        2 & 1273.06 & 24.38 & 139.64 & 110.62 & 800.64 & 26.26 & 17.18 & 0.62 & 46.86 & 106.54 & 0.32 \\
        3 & 603.26 & 12.38 & 70.70 & 74.50 & 389.78 & 10.38 & 5.76 & 0.46 & 10.70 & 28.16 & 0.44 \\
        4 & 274.80 & 7.04 & 38.46 & 36.36 & 171.14 & 4.08 & 2.34 & 0.24 & 3.76 & 11.26 & 0.12 \\
        5 & 132.64 & 4.82 & 20.14 & 16.36 & 81.52 & 2.52 & 1.32 & 0.12 & 1.56 & 4.26 & 0.04 \\
        6 & 68.58 & 3.12 & 12.16 & 7.68 & 40.50 & 1.38 & 0.84 & 0.04 & 0.86 & 1.96 & 0.02 \\
        7 & 40.34 & 2.04 & 7.90 & 4.24 & 23.20 & 0.86 & 0.54 & 0.04 & 0.44 & 1.06 & 0.02 \\
        8 & 25.86 & 1.50 & 5.38 & 2.80 & 14.40 & 0.54 & 0.34 & 0.02 & 0.26 & 0.64 & 0.02 \\
        9 & 17.26 & 0.98 & 3.72 & 1.96 & 9.48 & 0.36 & 0.24 & 0.02 & 0.14 & 0.38 & 0.02 \\
        10 & 12.22 & 0.76 & 2.74 & 1.38 & 6.60 & 0.24 & 0.14 & 0.00 & 0.12 & 0.26 & 0.00 \\
        11 & 8.92 & 0.54 & 2.00 & 1.00 & 4.86 & 0.16 & 0.10 & 0.00 & 0.06 & 0.18 & 0.00 \\
        12 & 6.52 & 0.44 & 1.50 & 0.72 & 3.46 & 0.12 & 0.08 & 0.00 & 0.04 & 0.12 & 0.00 \\
        13 & 4.94 & 0.32 & 1.16 & 0.56 & 2.62 & 0.08 & 0.06 & 0.00 & 0.04 & 0.10 & 0.00 \\
        14 & 3.96 & 0.26 & 0.94 & 0.44 & 2.08 & 0.06 & 0.04 & 0.00 & 0.04 & 0.08 & 0.00 \\
        15 & 3.02 & 0.20 & 0.72 & 0.34 & 1.56 & 0.06 & 0.04 & 0.00 & 0.04 & 0.06 & 0.00 \\
        16 & 2.44 & 0.18 & 0.58 & 0.28 & 1.26 & 0.04 & 0.04 & 0.00 & 0.00 & 0.04 & 0.00 \\
        17 & 2.02 & 0.16 & 0.48 & 0.24 & 1.04 & 0.04 & 0.02 & 0.00 & 0.00 & 0.04 & 0.00 \\
        18 & 1.58 & 0.12 & 0.40 & 0.18 & 0.80 & 0.02 & 0.02 & 0.00 & 0.00 & 0.02 & 0.00 \\
        19 & 1.26 & 0.10 & 0.30 & 0.14 & 0.64 & 0.02 & 0.02 & 0.00 & 0.00 & 0.02 & 0.00 \\
        20 & 1.06 & 0.08 & 0.28 & 0.12 & 0.54 & 0.02 & 0.02 & 0.00 & 0.00 & 0.02 & 0.00 \\
        \bottomrule
    \end{tabular}
\end{sidewaystable}

\newpage
\begin{figure}[H]
    \centering
    \begin{subfigure}[b]{0.49\textwidth}
        \centering
        \includegraphics[width=\textwidth]{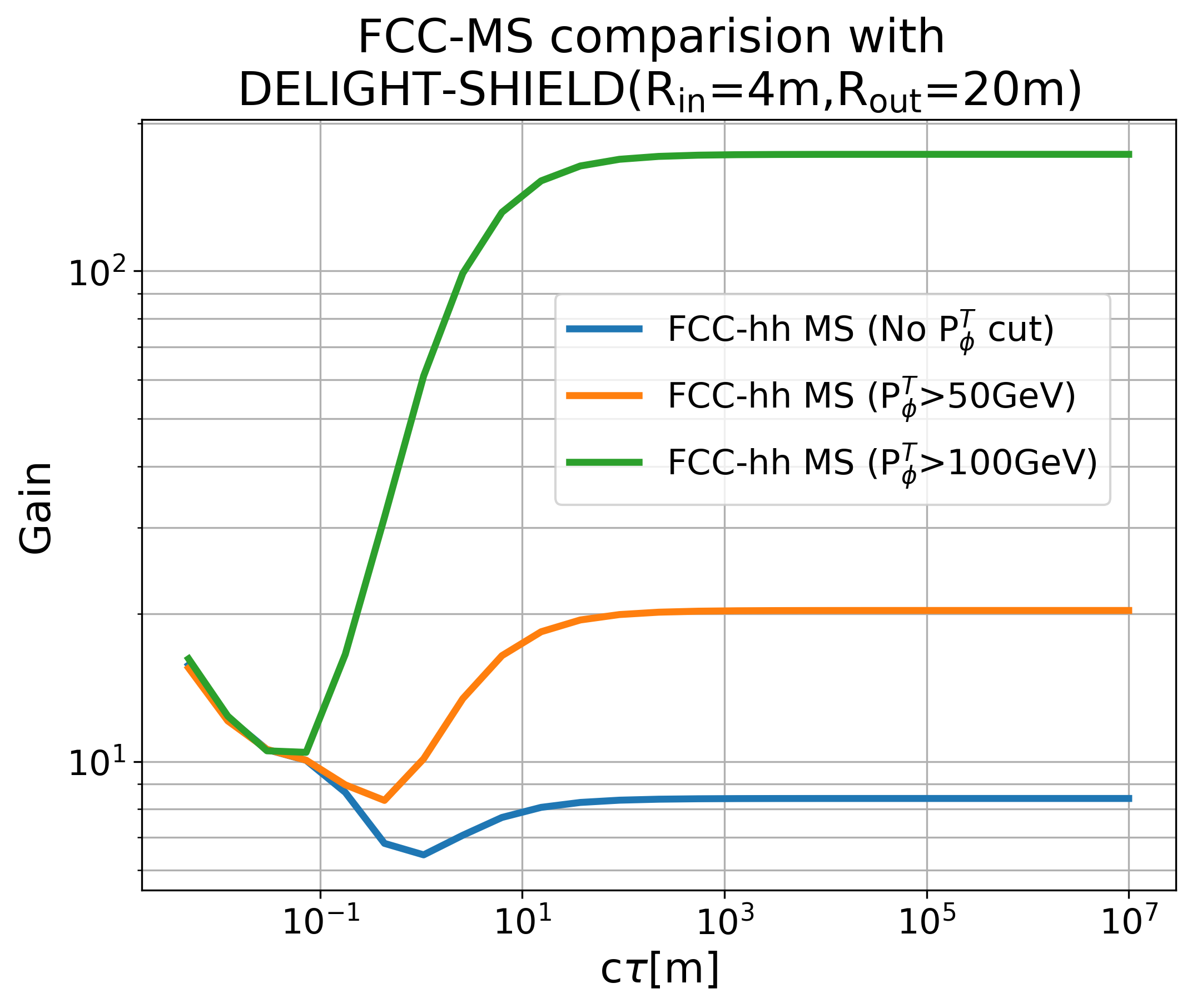}
        \caption{}
        \label{comarison1}
    \end{subfigure}
    \hfill
    \begin{subfigure}[b]{0.49\textwidth}
        \centering
        \includegraphics[width=\textwidth]{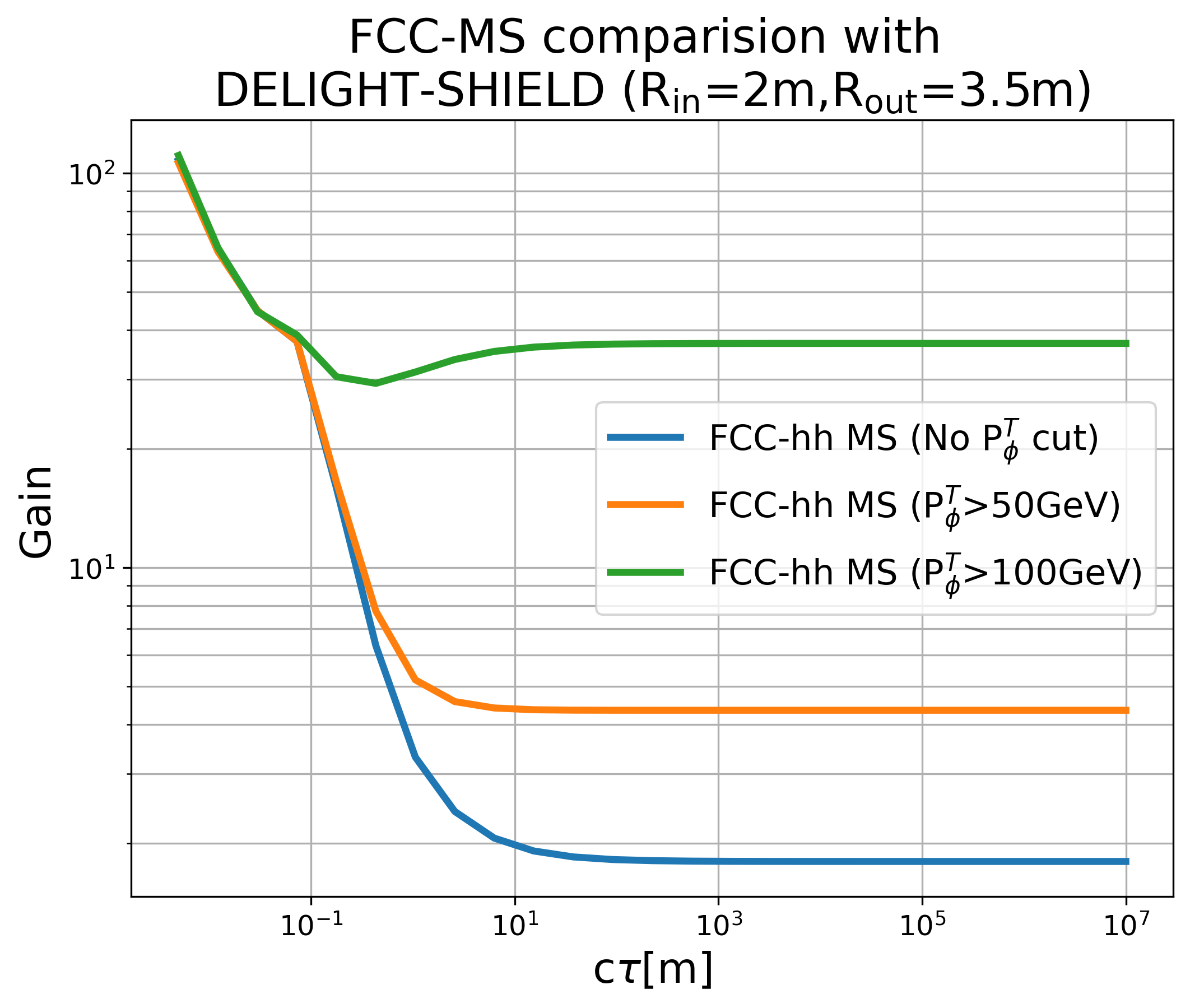}
        \caption{}
        \label{comarison1}
    \end{subfigure}
    \caption{Ratio of the sensitivity reach of \textsf{DELIGHT-SHIELD} 
    to that of the FCC-hh barrel muon spectrometer (MS) for 
    the benchmark process $h \rightarrow \phi\phi$ (where $m_{\phi}=6$\,GeV), as a 
    function of $c\tau$. Results are shown for two 
    representative geometric configurations of 
    \textsf{DELIGHT-SHIELD}($R_\text{in} = 4\,\text{m}$, 
    $R_\text{out} = 20\,\text{m}$ and $R_\text{in} = 2\,\text{m}$, 
    $R_\text{out} = 3.5\,\text{m}$), and for three 
    $p_\text{T}^{\phi}$ thresholds of 0, 50, and 
    $100\,\text{GeV}$ applied to the LLP signal in the muon 
    spectrometer. No $p_\text{T}$ requirement is applied to 
    \textsf{DELIGHT-SHIELD}. A ratio greater than unity indicates 
    improved sensitivity relative to the muon spectrometer.}
    \label{comparisons}
\end{figure}

We show the sensitivity gain, defined as the ratio of the \textsf{DELIGHT-SHIELD} reach to that of the FCC-hh muon spectrometer, in Fig.~\ref{comparisons} for the dark scalar having a mass of 6\,GeV, produced from Higgs boson decay.
For a large volume configuration  
with $R_\text{in} = 4\,\text{m}$ and $R_\text{out} = 20\,\text{m}$, 
the sensitivity gain is most pronounced at large $c\tau$, where the 
ratio saturates at values of approximately 8, 20, and 180 for 
no $p_\text{T}$ cut, a $50\,\text{GeV}$ cut, and a $100\,\text{GeV}$ 
cut, respectively. For a compact configuration with 
$R_\text{in} = 2\,\text{m}$ and $R_\text{out} = 3.5\,\text{m}$, 
the gain is instead most pronounced at small $c\tau$, saturating 
at values of approximately 2, 4, and 40 for large c$\tau$ under the same three 
$p_\text{T}$ scenarios. These results highlight the complementarity of different 
\textsf{DELIGHT-SHIELD} geometries and underline the substantial gain in 
sensitivity of \textsf{DELIGHT-SHIELD} over the FCC-hh muon spectrometer, 
which is a direct consequence of the significantly lower 
$p_\text{T}$ thresholds required for triggering in the shielded 
dedicated detector compared to the general-purpose muon spectrometer.

\begin{sidewaystable}[htbp]
    \centering
    \caption{Secondary particle yields per bunch crossing (Pileup=1000) for primary soft-QCD particles
             with $E > 0.5\,\text{GeV}$, for a thin shield (thickness of 10\,cm) at 100\,TeV.}
    \label{tab:particle_yield_summary_100TeV}
    \normalsize
    \setlength{\tabcolsep}{6pt}
    \begin{tabular}{c l l l l l l l l l l l}
        \toprule
        Energy[GeV] & $n_{tot}$ & $e/\gamma$ & $\pi^{\pm}$ & $p$ & $n$ & $K^{\pm}$ & $K^0$ & Nucleus & $\mu$ & $\nu$ & Hyperons \\ 
        \midrule
        1 & 16897.46 & 51.22 & 11357.48 & 1285.74 & 1410.72 & 1702.06 & 783.28 & 0.36 & 223.00 & 77.62 & 5.96 \\
        2 & 4815.12 & 28.84 & 2382.96 & 741.04 & 859.10 & 489.56 & 236.34 & 2.84 & 43.18 & 26.98 & 4.28 \\
        3 & 1699.30 & 16.36 & 747.98 & 327.44 & 353.90 & 156.48 & 78.48 & 1.54 & 9.30 & 5.66 & 2.16 \\
        4 & 630.82 & 10.66 & 282.54 & 116.70 & 124.16 & 58.30 & 32.22 & 0.74 & 2.88 & 1.56 & 1.06 \\
        5 & 264.42 & 7.38 & 123.60 & 43.16 & 46.84 & 25.86 & 15.44 & 0.22 & 1.10 & 0.46 & 0.36 \\
        6 & 127.98 & 4.80 & 63.18 & 17.80 & 19.74 & 12.92 & 8.48 & 0.02 & 0.58 & 0.22 & 0.22 \\
        7 & 70.88 & 3.20 & 35.48 & 9.40 & 10.40 & 7.04 & 4.78 & 0.02 & 0.32 & 0.10 & 0.12 \\
        8 & 41.86 & 2.22 & 21.12 & 5.36 & 6.00 & 4.04 & 2.78 & 0.00 & 0.18 & 0.06 & 0.10 \\
        9 & 26.96 & 1.58 & 13.74 & 3.44 & 3.84 & 2.44 & 1.74 & 0.00 & 0.10 & 0.02 & 0.08 \\
        10 & 17.90 & 1.14 & 9.26 & 2.22 & 2.42 & 1.58 & 1.14 & 0.00 & 0.08 & 0.02 & 0.04 \\
        11 & 12.64 & 0.88 & 6.48 & 1.58 & 1.72 & 1.10 & 0.80 & 0.00 & 0.04 & 0.00 & 0.04 \\
        12 & 8.86 & 0.66 & 4.56 & 1.08 & 1.18 & 0.74 & 0.56 & 0.00 & 0.02 & 0.00 & 0.02 \\
        13 & 6.48 & 0.52 & 3.38 & 0.76 & 0.82 & 0.52 & 0.40 & 0.00 & 0.02 & 0.00 & 0.02 \\
        14 & 5.04 & 0.44 & 2.62 & 0.60 & 0.64 & 0.38 & 0.32 & 0.00 & 0.02 & 0.00 & 0.02 \\
        15 & 3.70 & 0.32 & 1.90 & 0.42 & 0.46 & 0.30 & 0.24 & 0.00 & 0.02 & 0.00 & 0.02 \\
        \bottomrule
    \end{tabular}
\end{sidewaystable}

\begin{table}[htbp]
    \centering
    \caption{Particles spectra per bunch crossing from soft-QCD@14TeV ($|\eta| < 1.44$, $P_T>0.5$ GeV and $N_{PU}=200$)}
    \label{tab:softqcd_results_14TeV}
    \normalsize
    \setlength{\tabcolsep}{4pt}
    \begin{tabular}{l l l l l l l l l l}
        \toprule
        $E_{min}$ & $E_{max}$ & $e/\gamma$ & $\mu$ & $\pi^{\pm}$ & $K^{\pm}$ & $K_{L}$ & $K_{S}$ & $n$ & $p$ \\ 
        \midrule
        0.5 & 1.5 & 403.31 & 0.80 & 853.25 & 133.59 & 66.12 & 66.17 & 77.64 & 77.57 \\
        1.5 & 2.5 & 30.51 & 0.18 & 93.69 & 24.57 & 12.22 & 12.24 & 21.88 & 21.84 \\
        2.5 & 3.5 & 5.20 & 0.06 & 18.93 & 5.92 & 2.97 & 2.94 & 5.33 & 5.31 \\
        3.5 & 4.5 & 1.30 & 0.02 & 5.06 & 1.77 & 0.89 & 0.88 & 1.42 & 1.42 \\
        4.5 & 5.5 & 0.43 & 0.01 & 1.70 & 0.64 & 0.32 & 0.33 & 0.46 & 0.46 \\
        5.5 & 6.5 & 0.18 & 0.00 & 0.70 & 0.27 & 0.14 & 0.14 & 0.17 & 0.18 \\
        6.5 & 7.5 & 0.08 & 0.00 & 0.32 & 0.14 & 0.07 & 0.07 & 0.08 & 0.08 \\
        7.5 & 8.5 & 0.05 & 0.00 & 0.17 & 0.07 & 0.04 & 0.04 & 0.04 & 0.04 \\
        8.5 & 9.5 & 0.03 & 0.00 & 0.09 & 0.04 & 0.02 & 0.02 & 0.02 & 0.02 \\
        9.5 & 10.5 & 0.01 & 0.00 & 0.06 & 0.02 & 0.01 & 0.01 & 0.01 & 0.01 \\
        10.5 & 11.5 & 0.01 & 0.00 & 0.04 & 0.02 & 0.01 & 0.01 & 0.01 & 0.01 \\
        11.5 & 12.5 & 0.01 & 0.00 & 0.03 & 0.01 & 0.01 & 0.01 & 0.00 & 0.00 \\
        12.5 & 13.5 & 0.00 & 0.00 & 0.02 & 0.01 & 0.00 & 0.00 & 0.00 & 0.00 \\
        13.5 & 14.5 & 0.00 & 0.00 & 0.01 & 0.01 & 0.00 & 0.00 & 0.00 & 0.00 \\
        14.5 & 15.5 & 0.00 & 0.00 & 0.01 & 0.00 & 0.00 & 0.00 & 0.00 & 0.00 \\
        \bottomrule
    \end{tabular}
\end{table}

\begin{sidewaystable}[htbp]
    \centering
    \caption{Secondary particle yields per bunch crossing(Pileup=200) for primary soft-QCD particles
             with $E > 0.5\,\text{GeV}$, for a thin shield at 14\,TeV.}
    \label{tab:particle_yield_summary_14TeV}
    \normalsize
    \setlength{\tabcolsep}{6pt}
    \begin{tabular}{c l l l l l l l l l l l}
        \toprule
        Energy[GeV] & $n_{tot}$ & $e/\gamma$ & $\pi^{\pm}$ & $p$ & $n$ & $K^{\pm}$ & $K^0$ & Nucleus & $\mu$ & $\nu$ & Hyperons \\ 
        \midrule
        1 & 721.48 & 2.17 & 488.71 & 54.27 & 59.71 & 71.55 & 33.06 & 0.02 & 8.47 & 3.26 & 0.25 \\
        2 & 139.18 & 0.82 & 67.77 & 21.92 & 25.30 & 14.39 & 6.95 & 0.08 & 1.04 & 0.78 & 0.12 \\
        3 & 36.87 & 0.35 & 16.22 & 7.00 & 7.59 & 3.57 & 1.79 & 0.03 & 0.15 & 0.13 & 0.05 \\
        4 & 11.33 & 0.19 & 5.12 & 2.02 & 2.15 & 1.12 & 0.63 & 0.01 & 0.03 & 0.03 & 0.02 \\
        5 & 4.24 & 0.12 & 2.01 & 0.66 & 0.72 & 0.44 & 0.27 & 0.00 & 0.01 & 0.01 & 0.01 \\
        6 & 1.92 & 0.07 & 0.96 & 0.25 & 0.28 & 0.20 & 0.14 & 0.00 & 0.00 & 0.00 & 0.00 \\
        7 & 1.02 & 0.05 & 0.51 & 0.13 & 0.14 & 0.11 & 0.07 & 0.00 & 0.00 & 0.00 & 0.00 \\
        8 & 0.58 & 0.03 & 0.30 & 0.07 & 0.08 & 0.06 & 0.04 & 0.00 & 0.00 & 0.00 & 0.00 \\
        9 & 0.34 & 0.02 & 0.17 & 0.04 & 0.04 & 0.03 & 0.02 & 0.00 & 0.00 & 0.00 & 0.00 \\
        10 & 0.20 & 0.01 & 0.11 & 0.02 & 0.03 & 0.02 & 0.01 & 0.00 & 0.00 & 0.00 & 0.00 \\
        11 & 0.18 & 0.01 & 0.09 & 0.02 & 0.02 & 0.02 & 0.01 & 0.00 & 0.00 & 0.00 & 0.00 \\
        12 & 0.12 & 0.01 & 0.07 & 0.01 & 0.01 & 0.01 & 0.01 & 0.00 & 0.00 & 0.00 & 0.00 \\
        13 & 0.06 & 0.01 & 0.04 & 0.00 & 0.01 & 0.01 & 0.00 & 0.00 & 0.00 & 0.00 & 0.00 \\
        14 & 0.04 & 0.00 & 0.02 & 0.00 & 0.00 & 0.01 & 0.00 & 0.00 & 0.00 & 0.00 & 0.00 \\
        15 & 0.02 & 0.00 & 0.02 & 0.00 & 0.00 & 0.00 & 0.00 & 0.00 & 0.00 & 0.00 & 0.00 \\
        \bottomrule
    \end{tabular}
\end{sidewaystable}
\begin{figure}[htb!]
\centering
\includegraphics[width=0.7\linewidth]{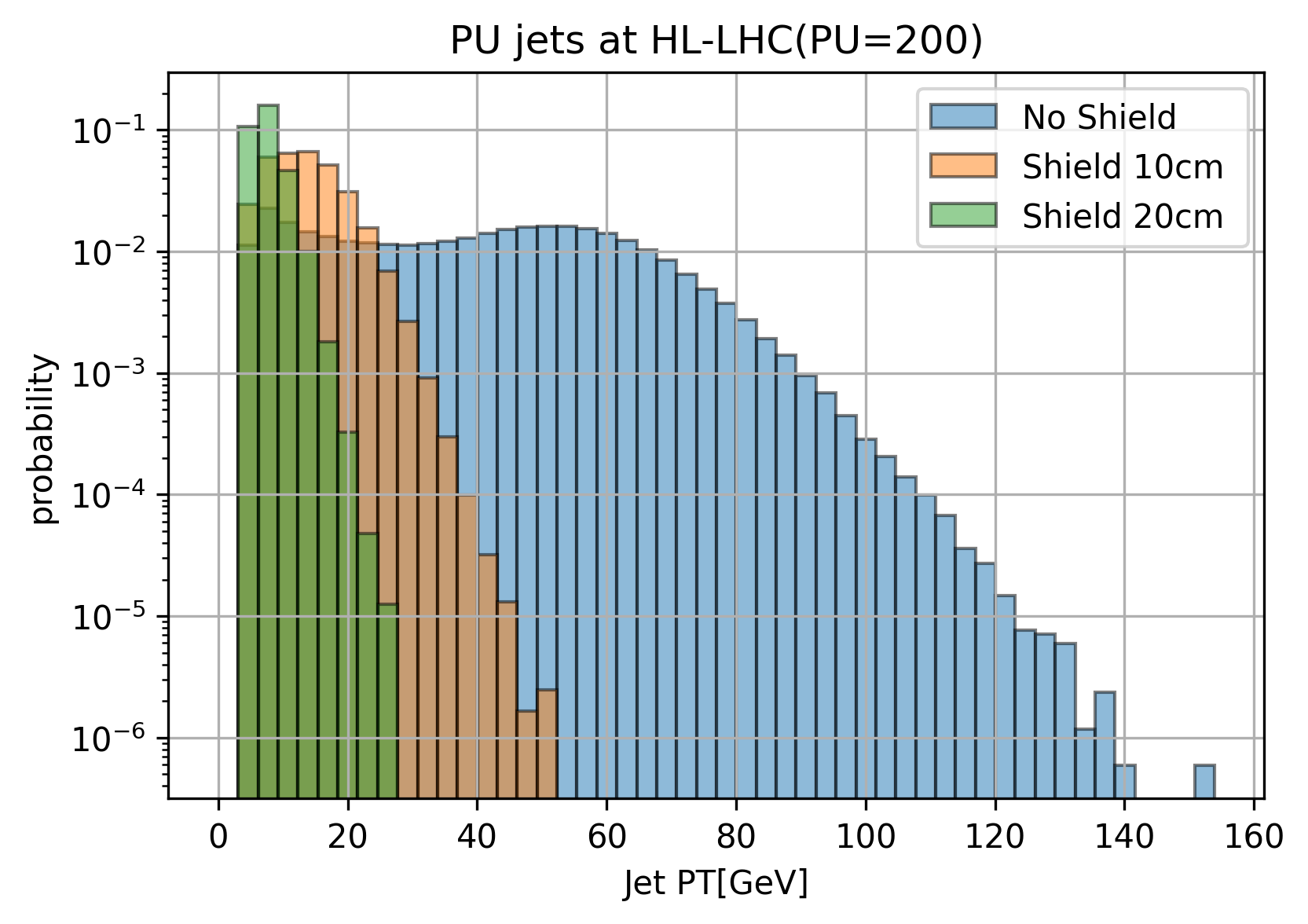}
\caption{Jet $p_\text{T}$ distribution for pile-up events 
at $\sqrt{s} = 14\,\text{TeV}$ with 
$\langle\mu\rangle = 200$, shown for three shielding 
configurations -- no shielding, $10\,\text{cm}$ of WCu80, 
and $20\,\text{cm}$ of WCu80. Jets are reconstructed 
using the anti-$k_t$ algorithm with $R = 0.4$ and 
$p_\text{T}^\text{jet} > 20\,\text{GeV}$.}
\label{jet_PT_hist} 
\end{figure}

\end{document}